\documentclass[onecolumn]{aastex631}

\usepackage{xcolor}

\usepackage{natbib} 
\usepackage{amsmath}
\usepackage{enumitem}
\shorttitle{Comet Bernardinelli-Bernstein}
\shortauthors{Bernardinelli et al.}

\defcitealias{Bernardinelli2019}{Paper I}
\defcitealias{y6tno}{B21}

\newcommand{\E}{\mathrm{E}}

\newcommand{\des}{\textit{DES}}
\newcommand{\au}{\,\mathrm{au}}

\newcommand\un{C/2014~UN$_{271}$}
\newcommand\unbb{\un\ (Bernardinelli-Bernstein)}
\newcommand\bb{BB}
\newcommand{\tensorG}{\ensuremath{\mathcal{G}}}
\newcommand{\xhat}{\ensuremath{\mathbf{\hat x}}}

\newcommand{\zhat}{\ensuremath{\mathbf{\hat z}}}
\newcommand{\ehat}{\ensuremath{\mathbf{\hat e}}}
\newcommand{\vhat}{\ensuremath{\mathbf{\hat v}}}

\newcommand{\vecx}{\ensuremath{\mathbf{x}}}
\newcommand{\vecL}{\ensuremath{\mathbf{L}}}
\newcommand{\vecr}{\ensuremath{\mathbf{r}}}
\newcommand{\vecb}{\ensuremath{\mathbf{b}}}
\newcommand{\vecv}{\ensuremath{\mathbf{v}}}
\newcommand{\gaia}{\textit{Gaia}}
\newcommand{\tess}{\textit{TESS}}
\newcommand{\afrho}{A\! f\! \rho}

\begin{document} 

\title{\unbb: the nearly spherical cow of comets}
\reportnum{For submission to ApJ Letters}
\reportnum{FERMILAB-PUB-21-449-AE}

\author[0000-0003-0743-9422]{Pedro H. Bernardinelli}
\affiliation{Department of Physics and Astronomy, University of Pennsylvania, Philadelphia, PA 19104, USA}
\email{pedrobe@sas.upenn.edu}

\author[0000-0002-8613-8259]{Gary M. Bernstein}
\affiliation{Department of Physics and Astronomy, University of Pennsylvania, Philadelphia, PA 19104, USA}
\email{garyb@physics.upenn.edu}

\author[0000-0001-7516-8308]{Benjamin~T.~Montet}
\affiliation{School of Physics, University of New South Wales, Sydney, NSW 2052, Australia}
\affiliation{UNSW Data Science Hub, University of New South Wales,
  Sydney, NSW 2052, Australia}

\author[0000-0002-0439-9341]{Robert Weryk}
\affiliation{Dept. of Physics \& Astronomy, University of Western Ontario, 1151 Richmond Street, London ON N6A 3K7, Canada}

\author{Richard Wainscoat}
\affiliation{Institute for Astronomy, University of Hawaii, 2680 Woodlawn Drive, Honolulu, HI 96822}

%
\author{M.~Aguena}
\affiliation{Laborat\'orio Interinstitucional de e-Astronomia - LIneA, Rua Gal. Jos\'e Cristino 77, Rio de Janeiro, RJ - 20921-400, Brazil}
\author{S.~Allam}
\affiliation{Fermi National Accelerator Laboratory, P. O. Box 500, Batavia, IL 60510, USA}
\author{F.~Andrade-Oliveira}
\affiliation{Instituto de F\'{i}sica Te\'orica, Universidade Estadual Paulista, S\~ao Paulo, Brazil}
\affiliation{Laborat\'orio Interinstitucional de e-Astronomia - LIneA, Rua Gal. Jos\'e Cristino 77, Rio de Janeiro, RJ - 20921-400, Brazil}
\author{J.~Annis}
\affiliation{Fermi National Accelerator Laboratory, P. O. Box 500, Batavia, IL 60510, USA}
\author{S.~Avila}
\affiliation{Instituto de Fisica Teorica UAM/CSIC, Universidad Autonoma de Madrid, 28049 Madrid, Spain}
\author{E.~Bertin}
\affiliation{CNRS, UMR 7095, Institut d'Astrophysique de Paris, F-75014, Paris, France}
\affiliation{Sorbonne Universit\'es, UPMC Univ Paris 06, UMR 7095, Institut d'Astrophysique de Paris, F-75014, Paris, France}
\author{D.~Brooks}
\affiliation{Department of Physics \& Astronomy, University College London, Gower Street, London, WC1E 6BT, UK}
\author{D.~L.~Burke}
\affiliation{Kavli Institute for Particle Astrophysics \& Cosmology, P. O. Box 2450, Stanford University, Stanford, CA 94305, USA}
\affiliation{SLAC National Accelerator Laboratory, Menlo Park, CA 94025, USA}
\author{A.~Carnero~Rosell}
\affiliation{Laborat\'orio Interinstitucional de e-Astronomia - LIneA, Rua Gal. Jos\'e Cristino 77, Rio de Janeiro, RJ - 20921-400, Brazil}
\author{M.~Carrasco~Kind}
\affiliation{Center for Astrophysical Surveys, National Center for Supercomputing Applications, 1205 West Clark St., Urbana, IL 61801, USA}
\affiliation{Department of Astronomy, University of Illinois at Urbana-Champaign, 1002 W. Green Street, Urbana, IL 61801, USA}
\author{J.~Carretero}
\affiliation{Institut de F\'{\i}sica d'Altes Energies (IFAE), The Barcelona Institute of Science and Technology, Campus UAB, 08193 Bellaterra (Barcelona) Spain}
\author{R.~Cawthon}
\affiliation{Physics Department, 2320 Chamberlin Hall, University of Wisconsin-Madison, 1150 University Avenue Madison, WI  53706-1390}
\author{C.~Conselice}
\affiliation{Jodrell Bank Center for Astrophysics, School of Physics and Astronomy, University of Manchester, Oxford Road, Manchester, M13 9PL, UK}
\affiliation{University of Nottingham, School of Physics and Astronomy, Nottingham NG7 2RD, UK}
\author{M.~Costanzi}
\affiliation{Astronomy Unit, Department of Physics, University of Trieste, via Tiepolo 11, I-34131 Trieste, Italy}
\affiliation{INAF-Osservatorio Astronomico di Trieste, via G. B. Tiepolo 11, I-34143 Trieste, Italy}
\affiliation{Institute for Fundamental Physics of the Universe, Via Beirut 2, 34014 Trieste, Italy}
\author{L.~N.~da Costa}
\affiliation{Laborat\'orio Interinstitucional de e-Astronomia - LIneA, Rua Gal. Jos\'e Cristino 77, Rio de Janeiro, RJ - 20921-400, Brazil}
\affiliation{Observat\'orio Nacional, Rua Gal. Jos\'e Cristino 77, Rio de Janeiro, RJ - 20921-400, Brazil}
\author{M.~E.~S.~Pereira}
\affiliation{Department of Physics, University of Michigan, Ann Arbor, MI 48109, USA}
\affiliation{Hamburger Sternwarte, Universit\"{a}t Hamburg, Gojenbergsweg 112, 21029 Hamburg, Germany}
\author{J.~De~Vicente}
\affiliation{Centro de Investigaciones Energ\'eticas, Medioambientales y Tecnol\'ogicas (CIEMAT), Madrid, Spain}
\author{H.~T.~Diehl}
\affiliation{Fermi National Accelerator Laboratory, P. O. Box 500, Batavia, IL 60510, USA}
\author{S.~Everett}
\affiliation{Santa Cruz Institute for Particle Physics, Santa Cruz, CA 95064, USA}
\author{I.~Ferrero}
\affiliation{Institute of Theoretical Astrophysics, University of Oslo. P.O. Box 1029 Blindern, NO-0315 Oslo, Norway}
\author{B.~Flaugher}
\affiliation{Fermi National Accelerator Laboratory, P. O. Box 500, Batavia, IL 60510, USA}
\author{J.~Frieman}
\affiliation{Fermi National Accelerator Laboratory, P. O. Box 500, Batavia, IL 60510, USA}
\affiliation{Kavli Institute for Cosmological Physics, University of Chicago, Chicago, IL 60637, USA}
\author{J.~Garc\'ia-Bellido}
\affiliation{Instituto de Fisica Teorica UAM/CSIC, Universidad Autonoma de Madrid, 28049 Madrid, Spain}
\author{E.~Gaztanaga}
\affiliation{Institut d'Estudis Espacials de Catalunya (IEEC), 08034 Barcelona, Spain}
\affiliation{Institute of Space Sciences (ICE, CSIC),  Campus UAB, Carrer de Can Magrans, s/n,  08193 Barcelona, Spain}
\author{D.~W.~Gerdes}
\affiliation{Department of Astronomy, University of Michigan, Ann Arbor, MI 48109, USA}
\affiliation{Department of Physics, University of Michigan, Ann Arbor, MI 48109, USA}
\author{D.~Gruen}
\affiliation{Faculty of Physics, Ludwig-Maximilians-Universit\"at, Scheinerstr. 1, 81679 Munich, Germany}
\author{R.~A.~Gruendl}
\affiliation{Center for Astrophysical Surveys, National Center for Supercomputing Applications, 1205 West Clark St., Urbana, IL 61801, USA}
\affiliation{Department of Astronomy, University of Illinois at Urbana-Champaign, 1002 W. Green Street, Urbana, IL 61801, USA}
\author{J.~Gschwend}
\affiliation{Laborat\'orio Interinstitucional de e-Astronomia - LIneA, Rua Gal. Jos\'e Cristino 77, Rio de Janeiro, RJ - 20921-400, Brazil}
\affiliation{Observat\'orio Nacional, Rua Gal. Jos\'e Cristino 77, Rio de Janeiro, RJ - 20921-400, Brazil}
\author{G.~Gutierrez}
\affiliation{Fermi National Accelerator Laboratory, P. O. Box 500, Batavia, IL 60510, USA}
\author{S.~R.~Hinton}
\affiliation{School of Mathematics and Physics, University of Queensland,  Brisbane, QLD 4072, Australia}
\author{D.~L.~Hollowood}
\affiliation{Santa Cruz Institute for Particle Physics, Santa Cruz, CA 95064, USA}
\author{K.~Honscheid}
\affiliation{Center for Cosmology and Astro-Particle Physics, The Ohio State University, Columbus, OH 43210, USA}
\affiliation{Department of Physics, The Ohio State University, Columbus, OH 43210, USA}
\author{D.~J.~James}
\affiliation{Center for Astrophysics $\vert$ Harvard \& Smithsonian, 60 Garden Street, Cambridge, MA 02138, USA}
\author{K.~Kuehn}
\affiliation{Australian Astronomical Optics, Macquarie University, North Ryde, NSW 2113, Australia}
\affiliation{Lowell Observatory, 1400 Mars Hill Rd, Flagstaff, AZ 86001, USA}
\author{N.~Kuropatkin}
\affiliation{Fermi National Accelerator Laboratory, P. O. Box 500, Batavia, IL 60510, USA}
\author{O.~Lahav}
\affiliation{Department of Physics \& Astronomy, University College London, Gower Street, London, WC1E 6BT, UK}
\author{M.~A.~G.~Maia}
\affiliation{Laborat\'orio Interinstitucional de e-Astronomia - LIneA, Rua Gal. Jos\'e Cristino 77, Rio de Janeiro, RJ - 20921-400, Brazil}
\affiliation{Observat\'orio Nacional, Rua Gal. Jos\'e Cristino 77, Rio de Janeiro, RJ - 20921-400, Brazil}
\author{J.~L.~Marshall}
\affiliation{George P. and Cynthia Woods Mitchell Institute for Fundamental Physics and Astronomy, and Department of Physics and Astronomy, Texas A\&M University, College Station, TX 77843,  USA}
\author{F.~Menanteau}
\affiliation{Center for Astrophysical Surveys, National Center for Supercomputing Applications, 1205 West Clark St., Urbana, IL 61801, USA}
\affiliation{Department of Astronomy, University of Illinois at Urbana-Champaign, 1002 W. Green Street, Urbana, IL 61801, USA}
\author{R.~Miquel}
\affiliation{Instituci\'o Catalana de Recerca i Estudis Avan\c{c}ats, E-08010 Barcelona, Spain}
\affiliation{Institut de F\'{\i}sica d'Altes Energies (IFAE), The Barcelona Institute of Science and Technology, Campus UAB, 08193 Bellaterra (Barcelona) Spain}
\author{R.~Morgan}
\affiliation{Physics Department, 2320 Chamberlin Hall, University of Wisconsin-Madison, 1150 University Avenue Madison, WI  53706-1390}
\author{R.~L.~C.~Ogando}
\affiliation{Observat\'orio Nacional, Rua Gal. Jos\'e Cristino 77, Rio de Janeiro, RJ - 20921-400, Brazil}
\author{F.~Paz-Chinch\'{o}n}
\affiliation{Center for Astrophysical Surveys, National Center for Supercomputing Applications, 1205 West Clark St., Urbana, IL 61801, USA}
\affiliation{Institute of Astronomy, University of Cambridge, Madingley Road, Cambridge CB3 0HA, UK}
\author{A.~Pieres}
\affiliation{Laborat\'orio Interinstitucional de e-Astronomia - LIneA, Rua Gal. Jos\'e Cristino 77, Rio de Janeiro, RJ - 20921-400, Brazil}
\affiliation{Observat\'orio Nacional, Rua Gal. Jos\'e Cristino 77, Rio de Janeiro, RJ - 20921-400, Brazil}
\author{A.~A.~Plazas~Malag\'on}
\affiliation{Department of Astrophysical Sciences, Princeton University, Peyton Hall, Princeton, NJ 08544, USA}
\author{M.~Rodriguez-Monroy}
\affiliation{Centro de Investigaciones Energ\'eticas, Medioambientales y Tecnol\'ogicas (CIEMAT), Madrid, Spain}
\author{A.~K.~Romer}
\affiliation{Department of Physics and Astronomy, Pevensey Building, University of Sussex, Brighton, BN1 9QH, UK}
\author{A.~Roodman}
\affiliation{Kavli Institute for Particle Astrophysics \& Cosmology, P. O. Box 2450, Stanford University, Stanford, CA 94305, USA}
\affiliation{SLAC National Accelerator Laboratory, Menlo Park, CA 94025, USA}
\author{E.~Sanchez}
\affiliation{Centro de Investigaciones Energ\'eticas, Medioambientales y Tecnol\'ogicas (CIEMAT), Madrid, Spain}
\author{M.~Schubnell}
\affiliation{Department of Physics, University of Michigan, Ann Arbor, MI 48109, USA}
\author{S.~Serrano}
\affiliation{Institut d'Estudis Espacials de Catalunya (IEEC), 08034 Barcelona, Spain}
\affiliation{Institute of Space Sciences (ICE, CSIC),  Campus UAB, Carrer de Can Magrans, s/n,  08193 Barcelona, Spain}
\author{I.~Sevilla-Noarbe}
\affiliation{Centro de Investigaciones Energ\'eticas, Medioambientales y Tecnol\'ogicas (CIEMAT), Madrid, Spain}
\author{M.~Smith}
\affiliation{School of Physics and Astronomy, University of Southampton,  Southampton, SO17 1BJ, UK}
\author{M.~Soares-Santos}
\affiliation{Department of Physics, University of Michigan, Ann Arbor, MI 48109, USA}
\author{E.~Suchyta}
\affiliation{Computer Science and Mathematics Division, Oak Ridge National Laboratory, Oak Ridge, TN 37831}
\author{M.~E.~C.~Swanson}
\affiliation{Center for Astrophysical Surveys, National Center for Supercomputing Applications, 1205 West Clark St., Urbana, IL 61801, USA}
\author{G.~Tarle}
\affiliation{Department of Physics, University of Michigan, Ann Arbor, MI 48109, USA}
\author{C.~To}
\affiliation{Department of Physics, Stanford University, 382 Via Pueblo Mall, Stanford, CA 94305, USA}
\affiliation{Kavli Institute for Particle Astrophysics \& Cosmology, P. O. Box 2450, Stanford University, Stanford, CA 94305, USA}
\affiliation{SLAC National Accelerator Laboratory, Menlo Park, CA 94025, USA}
\author{M.~A.~Troxel}
\affiliation{Department of Physics, Duke University Durham, NC 27708, USA}
\author{T.~N.~Varga}
\affiliation{Max Planck Institute for Extraterrestrial Physics, Giessenbachstrasse, 85748 Garching, Germany}
\affiliation{Universit\"ats-Sternwarte, Fakult\"at f\"ur Physik, Ludwig-Maximilians Universit\"at M\"unchen, Scheinerstr. 1, 81679 M\"unchen, Germany}
\author{A.~R.~Walker}
\affiliation{Cerro Tololo Inter-American Observatory, NSF's National Optical-Infrared Astronomy Research Laboratory, Casilla 603, La Serena, Chile}
\author{Y.~Zhang}
\affiliation{Fermi National Accelerator Laboratory, P. O. Box 500, Batavia, IL 60510, USA}
\collaboration{1000}{(The DES Collaboration)}
\suppressAffiliations

\begin{abstract}
  \unbb\ is a comet incoming from the Oort cloud which is remarkable
  in having the brightest (and presumably largest) nucleus of any
  well-measured comet, and having been discovered at heliocentric
  distance $r_h\approx29\au,$ farther than any Oort-cloud member.  We
  describe in this work the discovery process and observations, and
  the properties that can be inferred from images recorded until  the
  first reports of activity in June 2021.  The orbit has $i=95\degr,$
  with perihelion of 10.97~au to be 
  reached in 2031, and previous aphelion at $40,400\pm260\au.$
  Backwards integration of the orbit under a standard Galactic tidal
  model and known stellar encounters suggests this is a pristine new
  comet, with a perihelion of
$q\approx18\au$ on its previous perihelion passage 3.5~Myr ago.
The photometric data show
  an unresolved nucleus with absolute magnitude
  $H_r=8.0,$ colors that are typical of comet nuclei or Damocloids,
  and no secular trend as it traversed the range 34--23~au. 
For $r$-band geometric albedo $p_r,$ this implies a diameter of $150 (p_r/0.04)^{-0.5}$~km.
There is strong evidence of brightness fluctuations at $\pm0.2$~mag
level, but no rotation period can be discerned.  A coma,
nominally consistent with a ``stationary'' $1/\rho$ surface-brightness
distribution, grew in scattering cross-section at an exponential
rate from $\afrho\approx1$~m to $\approx150$~m as the comet approached from
28 to 20~au. The activity rate is consistent with a very simple model
of sublimation of a surface species in radiative equilibrium with the
Sun.  The inferred enthalpy
of sublimation matches those of $CO_2$ and
$NH_3$. More-volatile species such as $N_2,$ $CH_4,$ and $CO$
  must be far less abundant on the sublimating surfaces.
  \end{abstract}

\section{Introduction} 

Our knowledge of the content of the Oort cloud is highly
fragmentary---all inferences are based upon the small subset of its
members that are torqued into orbits with perihelia $q\lesssim10\au,$
and until recently only the subset of these which develop comae bright
enough to be noticed as comets.  The cometary activity makes the
objects easier to find and makes it easier to identify the composition
of the surface volatiles, but it can also obscure the properties of
the nuclear body.  The diversity of Oort cloud bodies has only
recently begun to be explored, with the discovery of objects having
varying levels of activity beyond the water frost line at
$\approx5\au$ \citep{meech08, Sarneczky,JewittK2HST,Jewitt21Activity,meech2017,Hui18,boattini}.  The discovery of \unbb\ (\bb\ hereafter,
for brevity) has expanded this known diversity substantially: as we
will elaborate below, it is probably the largest Oort body ever found
(indeed the largest of any kind of comet), and the first high-quality
observations were taken when \bb\ was at heliocentric distance
$r_h\approx29\au$ in 2014, well before the first announced detection of coma in
June 2021 at $r_h\approx20\au.$  In this work we will summarize the
observations in which \bb\ was discovered, and the inferences about
its composition and history that can be made from these and other
images taken until the recent first announcement of detectable activity.

\bb\ was discovered as part of the search for trans-Neptunian objects (TNOs) in the 80,000 exposures taken by the \textit{Dark Energy Survey} (\des) in the period 2013--2019 described fully in \citet{y6tno}. We refer to this paper for details of how $\approx108$~million single-night transient detections were identified and potential TNOs linked from amongst them.  The discovery of \bb\ was somewhat fortuitous because the search algorithms targeted objects at $r_h\ge29\au,$ while \bb\ was closer than this for all but its first \des\ exposures.  The \des\ search should therefore not be used to estimate the density of Oort-cloud members like \bb, though we can say that any object having $r_h>29\au$ and $m_r<23.8$ for $>2$~years of \des\ observing would have a high probability of detection.

\bb\ appears in 42 \des\ survey images in the $grizY$ filters on 25 distinct nights spanning 10 Oct 2014 to 26 Nov 2018.  Some of these images have artifacts that preclude precision photometry and/or astrometry, leaving 32 useful astrometric measures on 21 distinct nights, and 40 useful flux measures.  
The \textit{Solar System Object Image Search} service \citep{ssos}
finds additional archival imaging of \bb\ from \textit{WISE, CFHT, VST, VISTA} and \textit{PanSTARRS} observatories.  One \textit{VISTA} $z$-band ``pawprint'' from 20 October 2010 contains a measurable image of \bb, extending the arc and photometric record to $r_h=34.1\au.$  We measure positions and $gri$ fluxes of \bb\ in a series of 4 \textit{CFHT} exposures taken just before the first \des\ exposures, but do not attempt to measure the contemporaneous $u$-band exposure, which has only a marginal detection.  The object is not detectable in \textit{WISE} images taken during its primary mission in 2010 (E. Wright, private communication).   We did not attempt to recover \bb\ from the \textit{VST} images, even though some are previous to the \des\ epoch, since these have shorter exposures on a smaller telescope.  We also extract magnitudes from the \tess\ spacecraft imaging of the comet as it traversed Sector 3 in Sep--Oct 2018 and Sectors 29/30 in Aug--Oct 2020.
Circumstances, positions, fluxes, and uncertainties for \bb\ in these exposures are listed in Table~\ref{obstable}, with the \tess\ series each combined to a single mean flux.

Within 24 hours of publication of the \des\ discovery in MPEC 2021-M53
\citep{UN271}
on 19 June 2021, images were taken
showing visible coma \citep{CBET4989, LCO, MPEC2021-M83}. Analysis
of \tess\ data of BB indicated a large coma in 2018
\citep{TESS} and no detectable rotation period \citep{Ridden-Harper2021}.  In the
next section we examine the recent dynamics of \bb.
Section~\ref{sec:nuclear} examines the properties of the comet
nucleus, and Section~\ref{sec:coma} examines the onset of activity
before 2021.

\section{Astrometric properties}

\subsection{Measurements}
The \des\ astrometry is mapped to \gaia\ DR2 \citep{GaiaDR2}, using the
astrometric model presented in \cite{Bernstein2017astro}.  All
distortions due to the telescope, instrument and detections are known
to $\approx1$~mas RMS, and the color-dependent effects (differential
chromatic refraction in the atmosphere and lateral color distortions)
are corrected using the object's mean $g-i$ color. The position
uncertainties for the \des\ exposures (2014--2018) include the shot
noise from each detection as well as an anisotropic contribution from
the atmospheric turbulence \citep{Bernardinelli2019}. For the
\textit{VISTA} (2010) and \textit{CFHT} (2014) images, we retrieve
detrended images from the archives, remeasure the positions using
\texttt{SExtractor} windowed centroiding, and produce a polynomial
astrometric solution in the vicinity of \bb\ by referencing nearby
stars from the \gaia\ DR2 catalog.  Astrometry for
the \textit{PanSTARRS1} (2014--2019) exposures is extracted using
Gaussian fits to the comet and to field stars, with the latter
referenced to DR2.
One PS1 exposure is a $>3\sigma$ outlier from the orbit fit and is excluded from further consideration.

\subsection{Orbital properties and previous perihelion}
We determine the object's orbit using the method of \cite{Bernstein2000}, and we do not include non-gravitational forces in the orbit fit, as these have not been detected for \bb\ yet. The orbital elements and derived uncertainties are presented in Table \ref{orbtable}, and yield $\chi^2/\mathrm{dof} = 116.5/96$.  Considering only \des\ observations yields a consistent orbit with $\approx1.5\times$ larger uncertainties and $\chi^2/\mathrm{dof} = 66.4/58.$  The semi-major axis and inclination of the incoming orbit are 20,200~au and 95\fdg5, respectively, fully characteristic of Oort-cloud membership.  Perihelion of 10.95~au will be reached on 21 Jan 2031. The semi-major axis will be increased by 40\% after this perihelion.  

\begin{deluxetable}{ccccccc}
  \tablecaption{Osculating barycentric orbital elements}
  \label{orbtable}
\tablehead{\colhead{Epoch} & \colhead{$a$ (au)} & \colhead{$e$} & \colhead{$i$} & \colhead{$\Omega$} & \colhead{$\omega$} & \colhead{$t_{\rm peri}$ (JD)} }
\startdata
1950 & $20,200\pm130$ &  0.999458(4)  &  95\fdg4663 & 190\fdg0029 &  326\fdg2793(3) &  2462887.94(4) \\
2100 & $28,070\pm170$ &  0.999610(2)  &  95\fdg4606 & 190\fdg0093 &   326\fdg2438(4) &  2462887.87(9)
\enddata
\tablecomments{Elements are given at epochs before and after the
  current passage through the realm of the giant planets,} assuming only gravitational forces.  Uncertainties in the last digit are given in parentheses where they are sufficiently large.
\end{deluxetable}

It is of substantial interest to determine whether \bb\ has been appreciably warmed on previous perihelion passages.  
We study the past dynamics of \bb\ using a numerical procedure similar
procedure to that of \cite{krolikowska2018}, and also by analytic approximations.  We will consider perturbations by the Galactic tidal tensor \tensorG, assumed \citep[as in][]{heisler1986} to be diagonal in the Galactic frame rotating with the Sun where
\xhat\ points to the Galactic center and \zhat\ points to the North Galactic pole. This translates to a contribution to the Hamiltonian of the system in the form \citep{fouchard2004}
\begin{equation}
  \mathcal{H}_\mathrm{G} = \frac{1}{2}\vecx\cdot\tensorG\cdot\vecx = \mathcal{G}_1 \frac{x^2}{2} + \mathcal{G}_2 \frac{y^2}{2} + \mathcal{G}_3 \frac{z^2}{2}.
\end{equation}
We adopt the nominal Oort constants \citep{oort1927} $A = 15.1 \, \mathrm{km} \,\, \mathrm{s}^{-1} \, \mathrm{kpc}^{-1}$, $B = -13.4 \,\, \mathrm{km} \, \mathrm{s}^{-1} \, \mathrm{kpc}^{-1}$ from \cite{li2019} and a local stellar density $\rho_0 = 0.15 \,\, M_\sun \, \mathrm{pc}^{-3}$ from \cite{Vokrouhlicky2019}, so we have $\mathcal{G}_1 \equiv -(A-B)(3A+B)= -9.49 \times 10^{-16} \, \mathrm{yr}^{-2}$, $\mathcal{G}_2 \equiv (A-B)^2 = 8.48 \times 10^{-16} \, \mathrm{yr}^{-2}$ and $\mathcal{G}_3 \equiv 4\pi\mu\rho_0 - 2(B^2-A^2) = 8.59 \times 10^{-15} \, \mathrm{yr}^{-2}$. The angular velocity of the Sun is $\Omega_0 \equiv B - A = -28.5 \, \mathrm{km}\,\mathrm{s}^{-1}\,\mathrm{kpc}^{-1} = -2.91\times10^{-8} \, \mathrm{yr}^{-1}$.  

\begin{figure}[t]
  \plotone{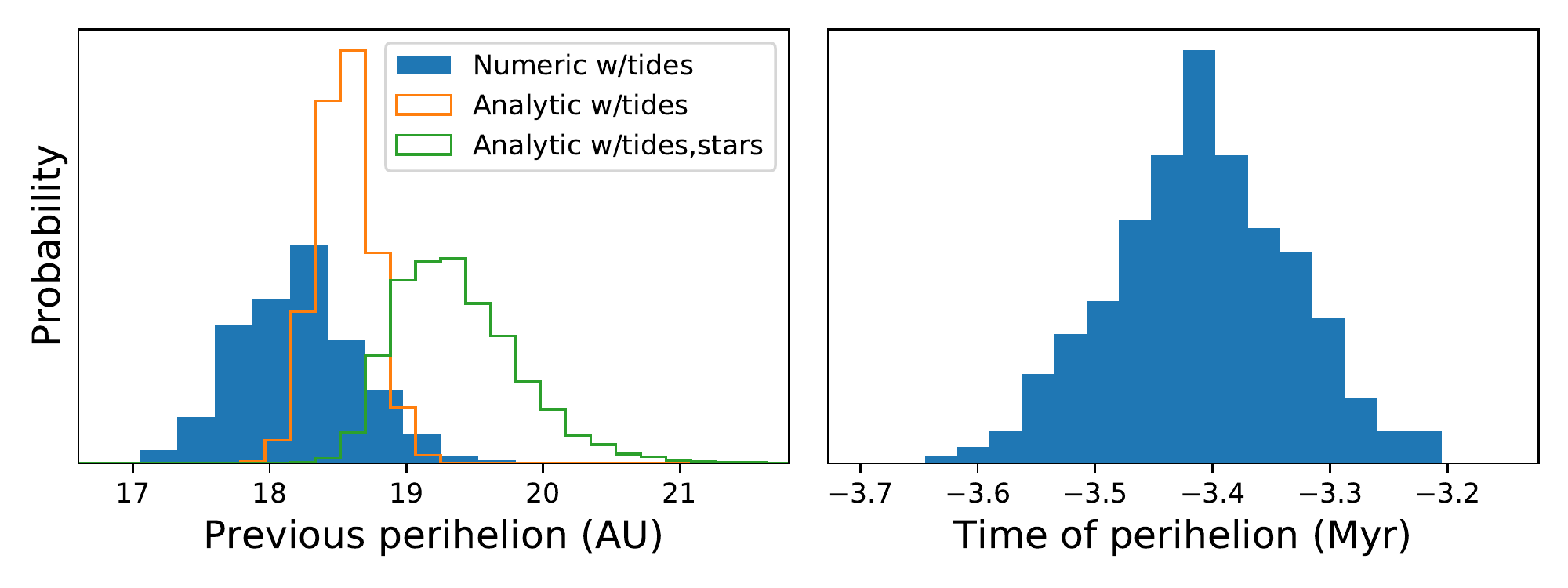}
  \caption{Distribution of properties of the previous perihelion of
    \bb.  The solid histograms show the predicted perihelion and its
    time of occurence in backward numerical integrations incorporating
    the giant planets and Galactic tides.  The orange and green open
    histograms show the result of analytic approximations that treat
    the solar system as a point mass, and use the Born approximation
    to a plunging comet orbit.  Both incorporate Galactic tides; the
    green histogram also includes the impulse approximation to the
    influence of 8 closely-approaching stars identified from the
    \gaia\ catalogs.  All plots marginalize over the uncertainties in
    the dynamical state of \bb\ and the stellar encounters. In all
    cases the previous perihelion is higher, at 17--21~au, than the
    current $q=11\au,$  and occurs $\approx3.4$~Myr ago.}
\label{im:previousperi}
\end{figure}

The numerical approach is to integrate, backwards in time, clones of the orbit solution sampled from the state vector covariance matrix.  We use the \textsc{WHFast} \citep{Wisdom1991,Rein2015} integrator of \textsc{REBOUND} \citep{rebound}, and include the giant planets as active perturbers as well as the effects of the Galactic tide using \textsc{REBOUNDx} \citep{tamayo2020}.  Figures~\ref{im:previousperi} show the histogram of previous-orbit perihelion distance and time from numerical integration of the sampled orbits, which are near $18.2\au$ and 3.41~Myr, respectively.

Analytically, we calculate the change in angular momentum
$\Delta\vecL$ imparted by the tidal torque over a full orbit in the
limit where the orbit is fully radial, $e\rightarrow 1.$  Near this
limit, $L^2=2kq,$ where $k$ is the barycentric gravitational constant
$1.0014GM_\odot,$  and $q$ is the perihelion.  We define \ehat\ as
the unit vector toward perihelion (inverse of aphelion direction).
The Born approximation then yields $\Delta \vecL =
5\pi\sqrt{\frac{a^7}{k}}\ehat \times (\tensorG\cdot\ehat).$  This
yields a previous perihelion of 18.3\,au for the nominal orbit, in
agreement with the numerical integration.  The ascending node
  of the previous passage has $r_h\approx20\au$ in the numerical integrations,
  leaving a $\sim1\%$ chance of an encounter within 2~au of Uranus.
  In the absence of such a perturbation, the perihelia of preceding
  orbits would have been increasingly higher.

We may also use the impulse approximation to assess the angular momentum imparted by passages of stars close to the Sun during the previous orbit.  For a star with mass $M_\star$ with closest approach to the Sun at point \vecb\ and velocity \vecv, while the comet is at position \vecr, the angular momentum imparted is
\begin{align}
  \Delta\vecL  &= \frac{2GM_\star}{v} \vecr \times \left[
                 \vecb \left( \frac{1}{|\vecb^\prime|^2} - \frac{1}{|\vecb|^2}\right) + \vhat \frac{\vecr\cdot\vhat}{|\vecb^\prime|^2}\right] \\
  \vecb^\prime & = \vecb - \vecr + \vhat(\vecr\cdot\vhat).
\end{align}
We use the list of reliable close stellar encounters ($b<1$~pc)
derived from the \gaia~DR2 catalog by \citet{bj18b}, restricted to
those with perihelion times $-4\,\textrm{Myr}<t_{ph}<0$ relative to
present.  We updated the stellar parameters for each star to the values
and uncertainties given in \gaia\ EDR3 \citep{edr3}.  This removes a very
strong perturber from this list (DR2 955098506408767360) and some
others, leaving 8 potential encounters over \bb's previous orbit
\citep[see also][]{bobylev2020}. We assume linear motion for the
perturbing stars and sample from the \gaia\ uncertainties. The net
effect of these encounters is to slightly decrease \vecL, i.e. to
raise the \emph{previous} perihelion by an amount that is well below
the effect of the Galactic tide. The left-hand panel of Figure~\ref{im:previousperi} shows the histograms of previous perihelion distance derived from the analytic approximations for the Galactic tide alone (in orange, sampling from \bb's orbital uncertainties) and for the Galactic tide plus stellar encounters (in green). 

The conclusion, which is robust to the details of the tidal model or
these 8 stars' dynamics, is that the previous passage of \bb\ was
\emph{further} from the Sun than the current one.  Indeed, under the
tidal model the perihelion has been getting smaller with each
successive passage for many orbits
into the past.  We conclude that \bb\ is a ``new'' comet in the
sense that there is no evidence for previous approach closer than
18~au to the Sun since ejection into the Oort cloud. Indeed, this may be the most pristine comet
  ever observed, in that we have detected it before it comes within
  Uranus's orbit, and it may never have done so on any previous orbit. It remains true, however, that our knowledge of stellar encounters is incomplete, and it is possible that some yet-unknown star's passage could have lifted \bb's perihelion from a lower value to its present one.

\section{Nuclear properties}
\label{sec:nuclear}

\subsection{Measurements}

We measure the flux in each \des\ image using scene-modeling
photometry, similar to \citet{Brout2019}. We define a target region
around each detection of $272\times272$ pixels (at 0\farcs264/pix), and simultaneously fit a
model for the object's flux and the background sources to all \des\
images from the same filter in this region of the sky. The background
is modeled as a grid of point sources that is present in all images,
while the object is modeled as a point source present in only the
detection image. Each point source is convolved with the point-spread
function (PSF, see \citealt{Jarvis2020} for a detailed description of
the \des\ PSF model) of each pixel location in each exposure. This
procedure also allows us to measure fluxes in exposures in which there
is no detection of the object, but the orbit indicates its
presence. Thus as seen in Table~\ref{obstable} there are \des\ images
having photometry but no useful astrometric data.  The resultant
fluxes and errors are rigorously correct for an unresolved image,
essentially using the central few arcseconds' signal, and thus
insensitive to any coma that does not have a central concentration.
The measures of diffuse flux in Section~\ref{sec:coma} confirm that
potential contamination of these point-source fluxes by coma flux will
be small for the \des\ data.  The same is not true for any of the
images taken after 2018---we will not use these in attempts to
characterize the nucleus, but will return to them when characterizing
the coma.
Flux
calibration for all \des\ exposures is determined to mmag precision as
described in \citet{DESDR2}. 

For the \textit{VISTA} and \textit{CFHT} detections, we acquire
detrended images from their respective
archives,\footnote{\url{http://archive.eso.org/wdb/wdb/adp/phase3_vircam/form}
  and \url{https://www.cadc-ccda.hia-iha.nrc-cnrc.gc.ca/en/cfht/}} and use \texttt{MAG\_AUTO} measurements from \texttt{SExtractor} \citep{Bertin1996}.  Each exposure is placed on the \des\ magnitude system by choosing a zeropoint to match the magnitudes found in the \des\ coadd catalogs in the corresponding filter for matching objects in field.  Bandpass differences between the  \textit{VISTA} $z$ and \textit{CFHT} $gri$ filters and their \des\ counterparts lead to color corrections that are well below the measurement errors on these points, and are ignored.  

The PS1 photometry given in Table~\ref{obstable} is derived by fitting
Gaussians to the comet images and to stars of known magnitude (the
$w$-band images use $r$-band magnitudes of the standards), and scaling
the Gaussian fits.  This photometry is less reliable and has lower
$S/N$ than \des\ data, so we will not make use of it in characterizing
the nucleus.  The PS1 measurements taken in 2019 are, however,
valuable for characterizing the development of coma between the end of
\des\ in 2018 and the 2021 recoveries.  We extract aperture photometry
for these images around the predicted positions of \bb\ to form the
curves of growth shown in Section~\ref{sec:coma}.  The magnitude
zeropoints of the $i$ and $w$ images are determined by comparison of
6\arcsec-diameter aperture photometry of bright stars to their $i$- and
$r$-band magnitudes in the \des\ catalogs.
The color terms between \textit{PS1} and \des\ bands are again well
below the measurement errors \citep[Eq.~B6 of][]{DESDR2}. 

\tess\ observations consist of ``sectors,'' $24 \times 96$ degree
regions of the sky observed nearly continuously for approximately four
weeks \citep{Ricker15}.  In its survey of the southern sky, \tess\
observed \bb\ in three sectors, one in late 2018 and two in
late 2020.  For all three sectors, we identify and cut out an
approximately $1.5 \times 0.5$ degree region of the \tess\ full-frame
images (FFIs) along the path of the comet with the \texttt{tesscut}
tool \citep{tesscut19}.  We then apply a difference imaging scheme
aimed towards removing background stars by, for each frame and each
pixel, subtracting the mean flux observed in that pixel in all
cadences observed between 5 and 10 hours from the time of the frame of
interest.

At each frame, we then measure the flux of the target in an aperture
of $5\times5$ of \tess's 21\arcsec\ pixels.  We apply the same
method to nearby stars on the detector with low ($<1\%$) levels of
photometric variability and well-characterized \tess\ 
magnitudes to transform our measured fluxes to magnitudes.  The
scatter of the residuals for stars on this scale is 0.15 magnitudes,
likely due to crowding of faint stars and intrapixel sensitivity
variations on the \tess\ detector \citep{Vorobiev19}. These issues
should be less dramatic for \bb\ due to its motion across the
detector, nonetheless we apply this 0.15 magnitude uncertainty
conservatively on the individual magnitudes.
The
brightening of $2.01\pm0.04$~mag between the two \tess\ epochs 
is more reliably determined than the magnitude at either epoch. 
For an object of solar color, we should find $r-T =
r_\odot-T_\odot=4.61-4.26=0.35$~mag \citep{willmer,tesszpt}.  

\begin{figure}
\centering
\includegraphics[width=0.7\textwidth]{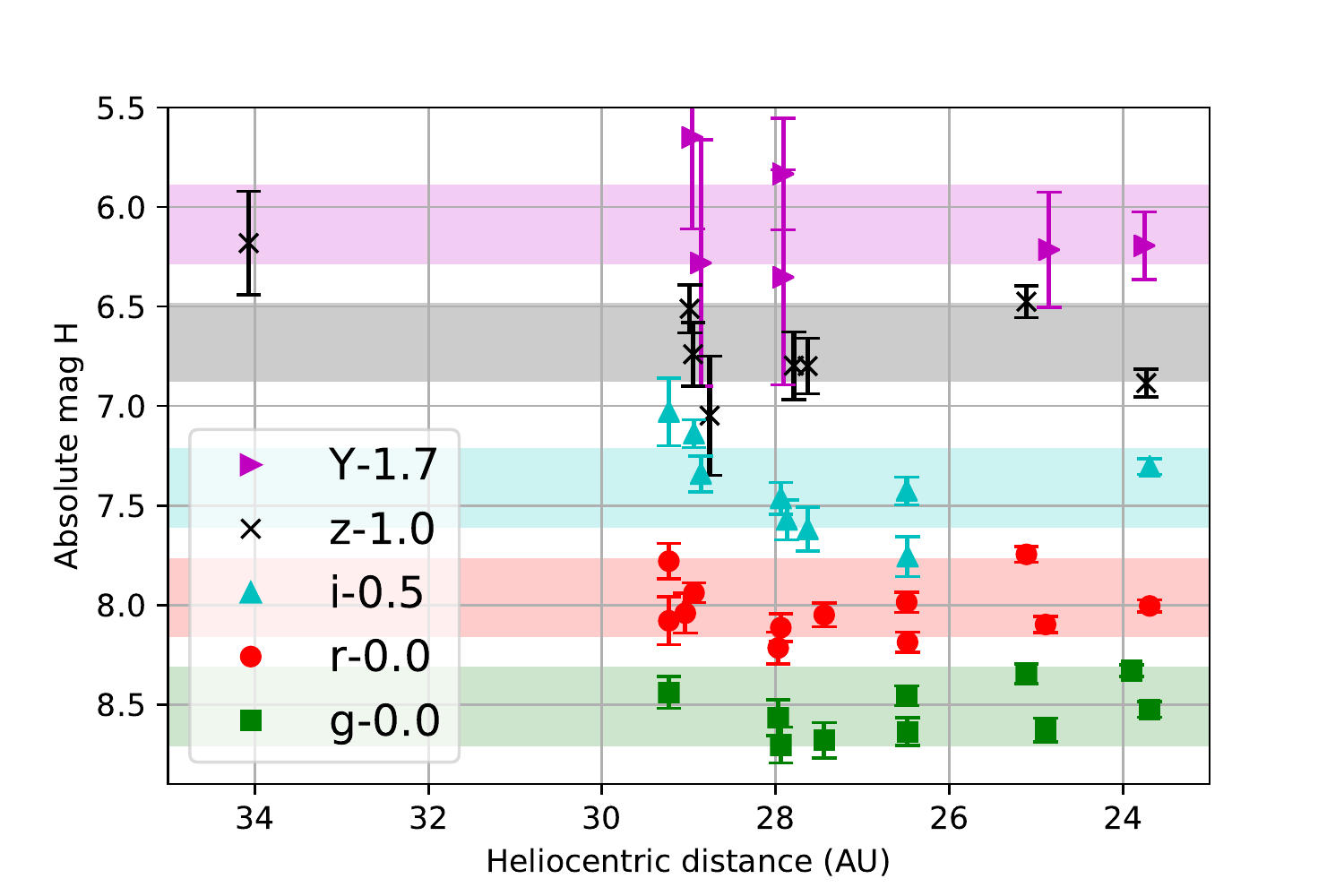}
\caption{Photometry of \bb\ is plotted vs heliocentric distance
  (bright is up, time runs to the right).  This combines observations
  from \des, \textit{CFHT}, and \textit{VISTA}. For clarity we have
  shifted the $i, z,$ and $Y$ data by the amounts noted in the
  legend. The horizontal colored bands are centered on the mean $H_b$
determined for each band $b$, and have the width of the best-estimate
$\pm0.20$~mag of light-curve variation. } 
\label{im:allphot}
\end{figure}

\subsection{Color, variability, and size}
\begin{figure}
\plottwo{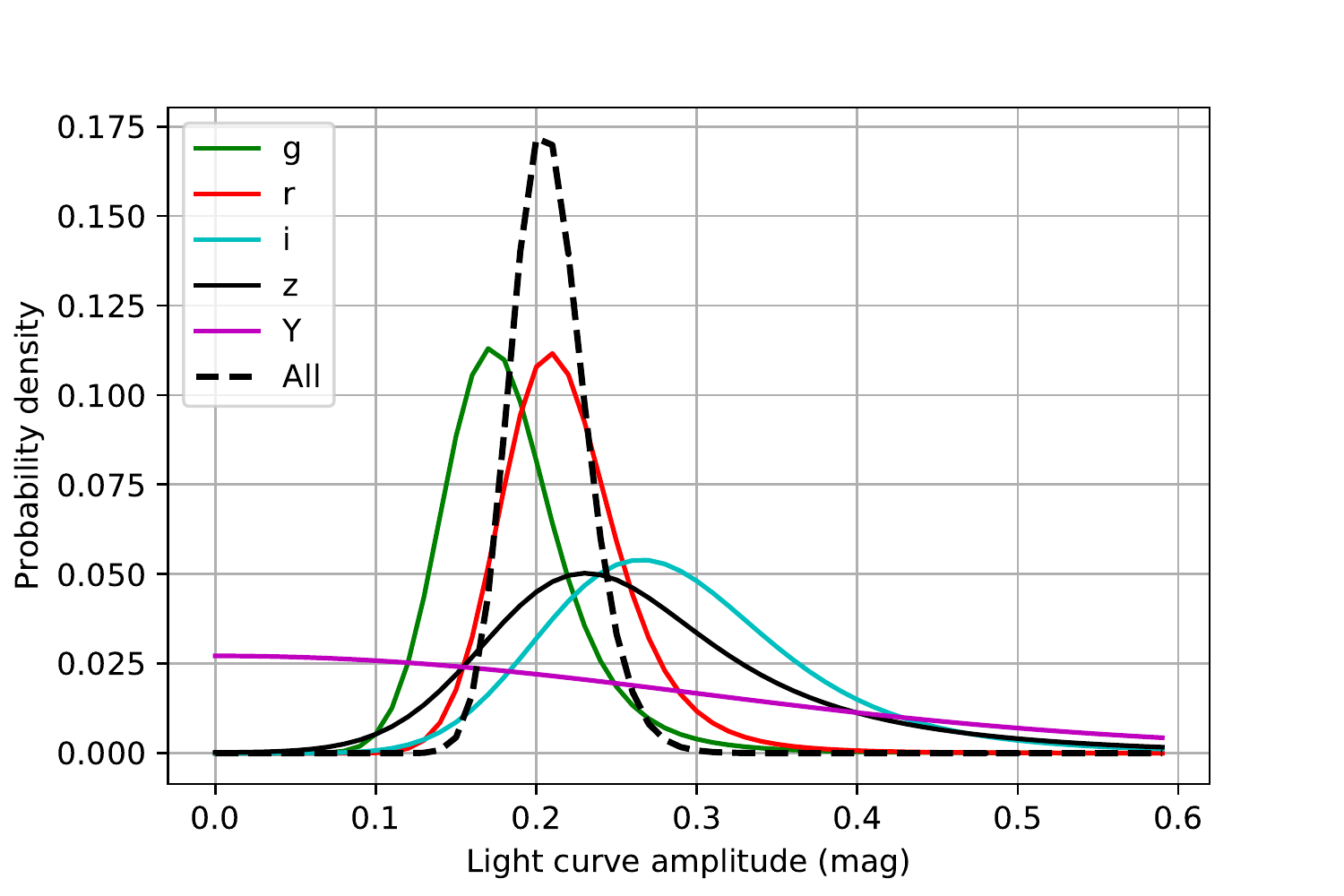}{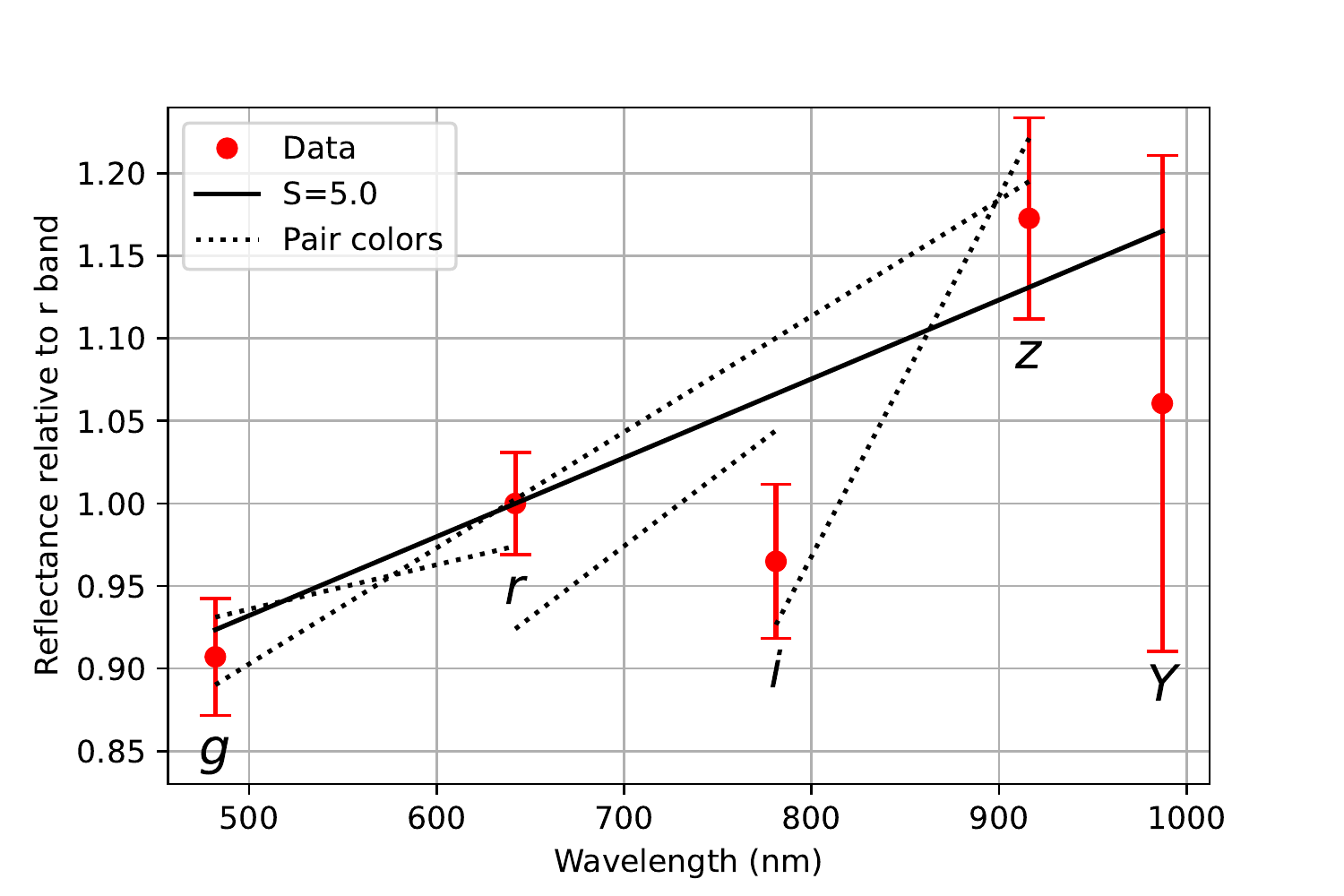}
\caption{\textit{Left:} Posterior probability of the amplitude $A$ of the comet
  light curve variations $\Delta H = A\sin\phi$ as derived from 2018 and earlier.
  The individual bands' constraints are consistent, and the combined
  result of $A=0.20\pm0.03$~mag strongly excludes a constant absolute
  magnitude.
  \textit{Right:} The relative surface reflectance of \bb,
  normalized to the nominal $r$-band value, is plotted vs wavelength.
  The symbols are derived from the mean $H$ values found from
  maximizing the probability in Eq.~(\ref{eq:prob}).  The dotted line
  segments show the reflectance slopes implied by pairs of
  close-in-time exposures---the vertical positioning of these is arbitrary.  The solid line is a model of linear
  dependence of reflectance on wavelength, with a slope of $S=5\%$ per
  100~nm.  This is similar to, but slightly more neutral than, colors
  reported for other long-period-comet nuclei.} 
\label{im:lcamp}
\end{figure}

We fit all of the valid photometry from \textit{VISTA, CFHT,} and
\des\ to a model in which there is a fixed absolute magnitude $H_b$ in
each band $b$, and an achromatic light curve fluctuation $\Delta H =
A\sin\phi.$ These data are plotted in Figure~\ref{im:allphot}.  The
illumination phase is between 1\fdg4 and 2\fdg5 for all observations
here, so we ignore phase terms in converting observed magnitudes to
$H$.  We have insufficient data to determine the light-curve phases $\phi_i$ for
each exposure, so we consider each observation to have a random,
independent $\phi_i\in[0,2\pi].$  The posterior probability of the light
curve amplitude and the ``true'' $H_b$, given observations of $H_i$ in
band $b_i$ with uncertainty $\sigma_i$ for each observation $i$, is 
\begin{equation}
  p\left( A,\{H_b\} | \{H_i, \sigma_i\}\right) \propto
  \prod_{i\in \textit{obs}} \int \mathrm{d}\phi \, \exp\left[-(H_i-H_{b_i}-A \sin \phi)^2 / 2\sigma_i^2\right].
  \label{eq:prob}
\end{equation}
Figure~\ref{im:lcamp} (left) plots the posterior probability for the
light-curve semi-amplitude $A.$  All of the bands are consistent and
combine to yield  $A=0.20\pm0.03$~mag.   This is a strong detection of
variability in excess of the measurement errors. From
Figure~\ref{im:allphot} it is clear that most of this variability is
in short-term variation, not a long-term trend, consistent with a
nuclear body with 10--20\% departures from sphericity.  \citet{tessLC}
report a non-detection of variation in the \tess\ photometric
time series, though no upper limit is reported.  The \tess\ 
photometry has the majority of its flux coming from the coma, which
will suppress the amplitude of any nuclear light curve.

The fitting process yields estimates of the mean absolute magnitude of
the comet of $H=\{8.51\pm0.04, 7.96\pm0.03, 7.91\pm0.05, 7.68\pm0.06,
7.79\pm0.14\}$ in the $grizY$ bands, respectively.
Figure~\ref{im:lcamp} (right) plots the implied reflectivity in each
band, normalized to unity in the $r$ band, showing a color only
slightly redder than neutral, perhaps even slightly blue in $r-i$.  An
alternative method of deriving colors for \bb\ is to find pairs of
exposures taken in different bands within 5--10 minutes of each other,
so that light-curve variations are unimportant.  This results in color
estimates of $g-r=0.49\pm0.01$; $r-i=0.22\pm0.02$; $i-z=0.32\pm0.09$;
and $g-z=0.87\pm0.04.$  These agree with the mean-$H$ method, except
that the pair-based $r-i$ color is significantly redder. 

\citet{Jewitt2015} presents colors for various outer-solar-system
bodies, quantified by a fit to a model of linear reflectance vs
wavelength with slope of  $S\%$ per 100~nm when normalized to
unit reflectivity at 550~nm ($V$-band). The solid line in the right panel of
Figure~\ref{im:lcamp} shows that $S=5$ approximates the data for \bb,
though with a potential absorptive feature in the
$i$-band. \citet{Jewitt2015} reports that potential relatives of \bb,
namely long-period comet (LPC) nuclei and Damocloids, have typical $S$ values
of 10 and 15, respectively.  These Oort bodies are
  significantly bluer than the TNO populations. Comet \bb\ shares this
deviation from the TNO colors, in fact appearing a bit more neutral 
than the few other well-measured Oort-cloud migrants. 

\begin{figure}
  \plottwo{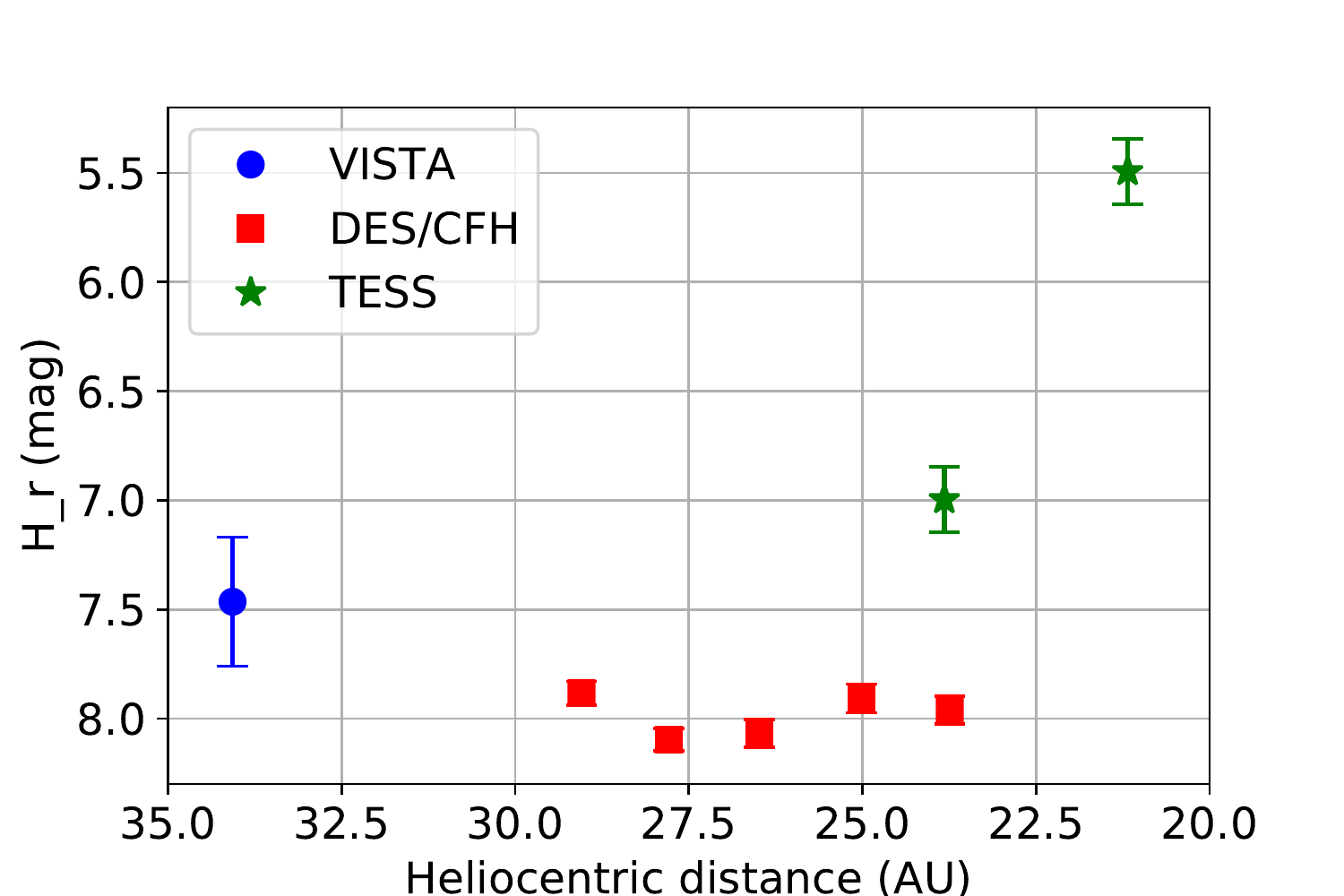}{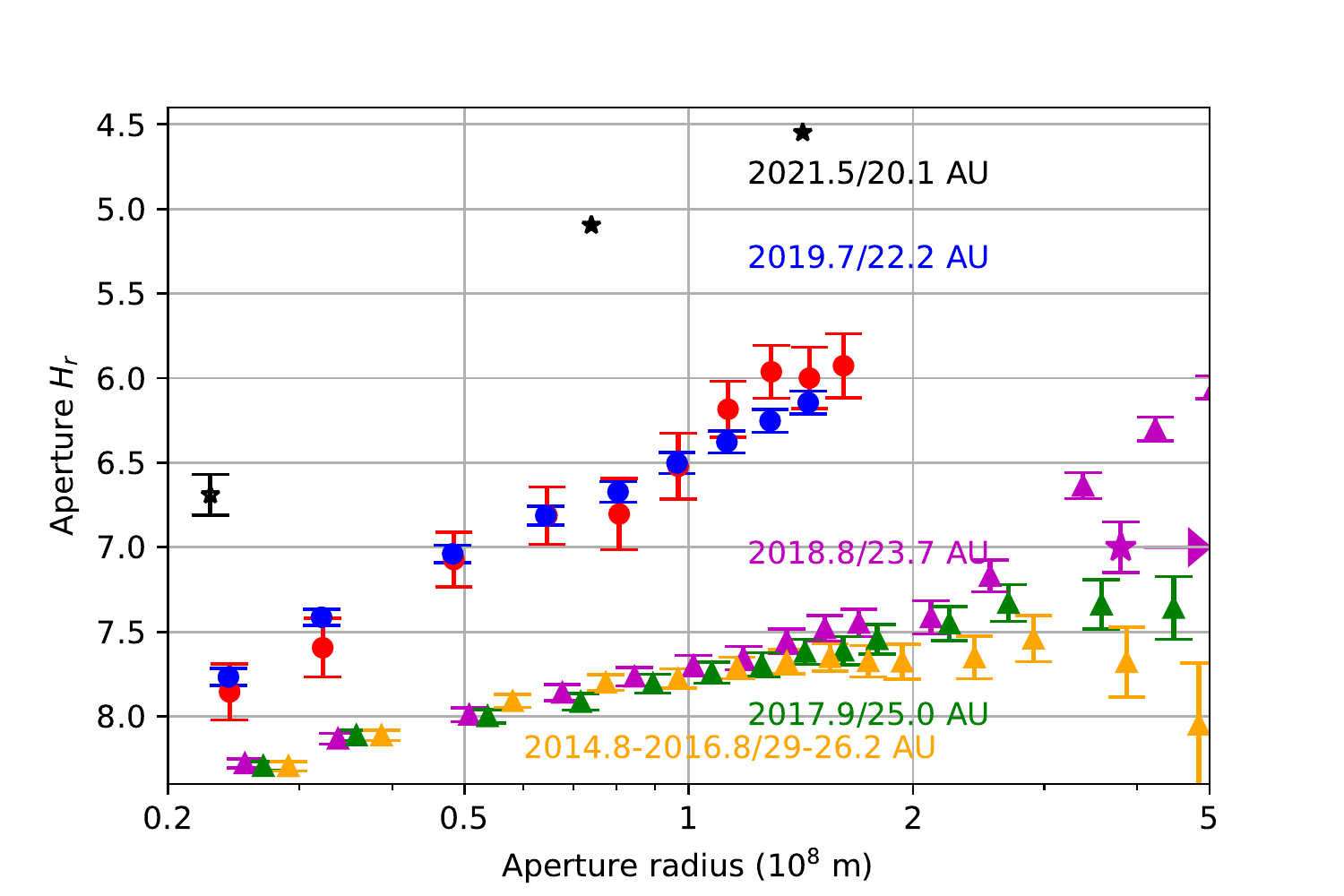}
  \caption{\textit{Left:} The annual average absolute magnitudes, transformed to $r$
    band using the measured colors, are plotted vs heliocentric
    distance (bright is up, time advances to the right).  The
    ground-based data are consistent with no overall brightening
    during the approach from 34 to 22~au. The \tess\ data, however, show a highly
    significant brightening of 1.5~mag between 23 and 21~au (2018 to
    2020).  Furthermore the earlier \tess\ epoch shows
    significantly higher flux than the contemporaneous \des\
    measurements, suggesting the presence of a very diffuse coma at
    this time.
    \textit{Right:} Curves of growth of \bb's absolute magnitude vs physical
    aperture radius are shown, with the epoch and $r_h$ as labeled.  From the bottom up:
    the triangles are from \des\ data, divided into three time periods
    as labeled.  The magenta star is the TESS observation during
    the final \des\ season, which has aperture radius $\approx10^9$~m.
  The blue and red circles are from \textit{PS1}
    observations in the $w$ and $i$ bands, respectively, on two
    different nights of August 2019;  The black stars are aperture
    data taken in June 2021 by \citet{LCO} (with error bar) and
    \citet{dekelver} (no uncertainties specified).  In all cases the
    curve of growth of stellar sources is flat at radii
    $\gtrsim7\times10^7$~m. The presence of activity is detectable
  at large radii in images as early as 2017.  The rise of the 2018
  curve beyond $3\times10^8$~m could be an artifact of sky
  subtraction, but the curve is consistent with the $\approx1$~mag
  difference between contemporaneous \tess\ and \des\ in the left panel.}
\label{im:secular}
\end{figure}

If the flux measurements in these $\le2018$
exposures were significantly contaminated with coma rather than being
predominately nuclear, we might expect the measured $H$ to increase as
the comet approaches the Sun.  In Figure~\ref{im:secular}, we present
the $H_r$ averaged over all photometric observations from a given
season.  Exposures from band $b$ are shifted to $r$ band using the
$H_r-H_b$ from the posterior maximization above.  An RMS error of
$0.20/\sqrt{2}$~mag is added in quadrature to each measurement
error to include noise from random sampling of a sinusoidal light
curve.  While the year-to-year means of the \des\ observations are
formally inconsistent with a constant magnitude, the potential
year-scale variation is small ($\approx0.1$~mag) and shows no long
term trend.  Indeed the 2010 \textit{VISTA} 
observation is consistent with a constant magnitude as well, so
there is no evidence for a brightening of \bb's absolute magnitude as
it moves from $r_h=34.1$ to 23.7~au for apertures of $\approx1\arcsec$ size.

Under the assumption that the $H_r=7.96$ derived above is entirely
from a spherical nucleus with geometric albedo in the $r$ band of
$p_r$ and density $\rho,$ the diameter, mass, and escape velocity of \bb\ are
\begin{align}
  D & = \left(\frac{p_r}{0.04}\right)^{-1/2} \, 155\,\textrm{km}
  \label{nominaldiam}\\
  m & = \left(\frac{p_r}{0.04}\right)^{-3/2} \left(\frac{\rho}{1\,\textrm{g}\, \textrm{cm}^{-3}}\right)\, 
      \, 2.0\times10^{18}\,\textrm{kg} \label{nominalmass}\\
  v_{\rm esc} & =  \left(\frac{p_r}{0.04}\right)^{-1/2} \left(\frac{\rho}{1\,\textrm{g}\, \textrm{cm}^{-3}}\right)^{1/2} \,
  58\,\textrm{m}\,\textrm{s}^{-1}.\label{nominalvesc}
\end{align}
At the nominal assumed albedo, this makes \bb\ a factor of 2.5 larger
in diameter than C/1995~O1 (Hale-Bopp) \citep{Fernandez}, another
LPC that is the largest of any comet in the past century
\citep{Lamy2004},
and had $H_r\approx9.7$ at incoming $r_h=6.4\au$
  \citep{szabo}.

\section{Coma development}
\label{sec:coma}

The \tess\ photometry, plotted as stars in the left-hand panel of
Figure~\ref{im:secular}, shows a definitive 1.5~mag increase in $H$
during the two-year journey from $r_h=23.8\au$ to 21.2~au, after the
\des\ observations end, from which we infer an increase in activity
before the June 2021 discovery of the coma at $r_h=20.2\au.$  More
surprisingly, the \tess\ images from 2018 show a substantially
brighter $H$ than the \des\ photometry at the same time period, by
$\approx1.0\pm0.15$~mag.
Furthermore, \citet{TESS}
reports that the 2018 \tess\ images are resolved, with a
Gaussian fit yielding $\textrm{FWHM}\approx2.92$ of the 21\arcsec\
pixels, while unresolved sources are $\le2.06$~pix.  A simple
quadrature subtraction suggests that the intrinsic comet angular size
is at least 2.06~pix$=43\arcsec,$ e.g.\ a Gaussian with
$\sigma>18\arcsec.$  A coma of this size would have gone undetected in
the \des\ scene-modelling photometry \emph{if the coma did not have a strong central
  concentration in the inner 1--2\arcsec.}

With this in mind we re-analyze the \des\ images and the \textit{PS1}
images from 2019 for signs of 10\arcsec-scale emission.
Figure~\ref{im:secular} (right) shows the curves of growth derived from aperture
photometry of these images, as well as June~2021 observations reported
by \citet{LCO} and \citet{dekelver}.  It is apparent that the coma was
already present in the \textit{PS1} images at $r_h=22.6\au,$ indeed
also for most of the \des\ observations, albeit not at a level that
precludes our attribution of the PSF-fitting fluxes to the nucleus.

\begin{figure}
  \centering
  \includegraphics[width=0.6\textwidth]{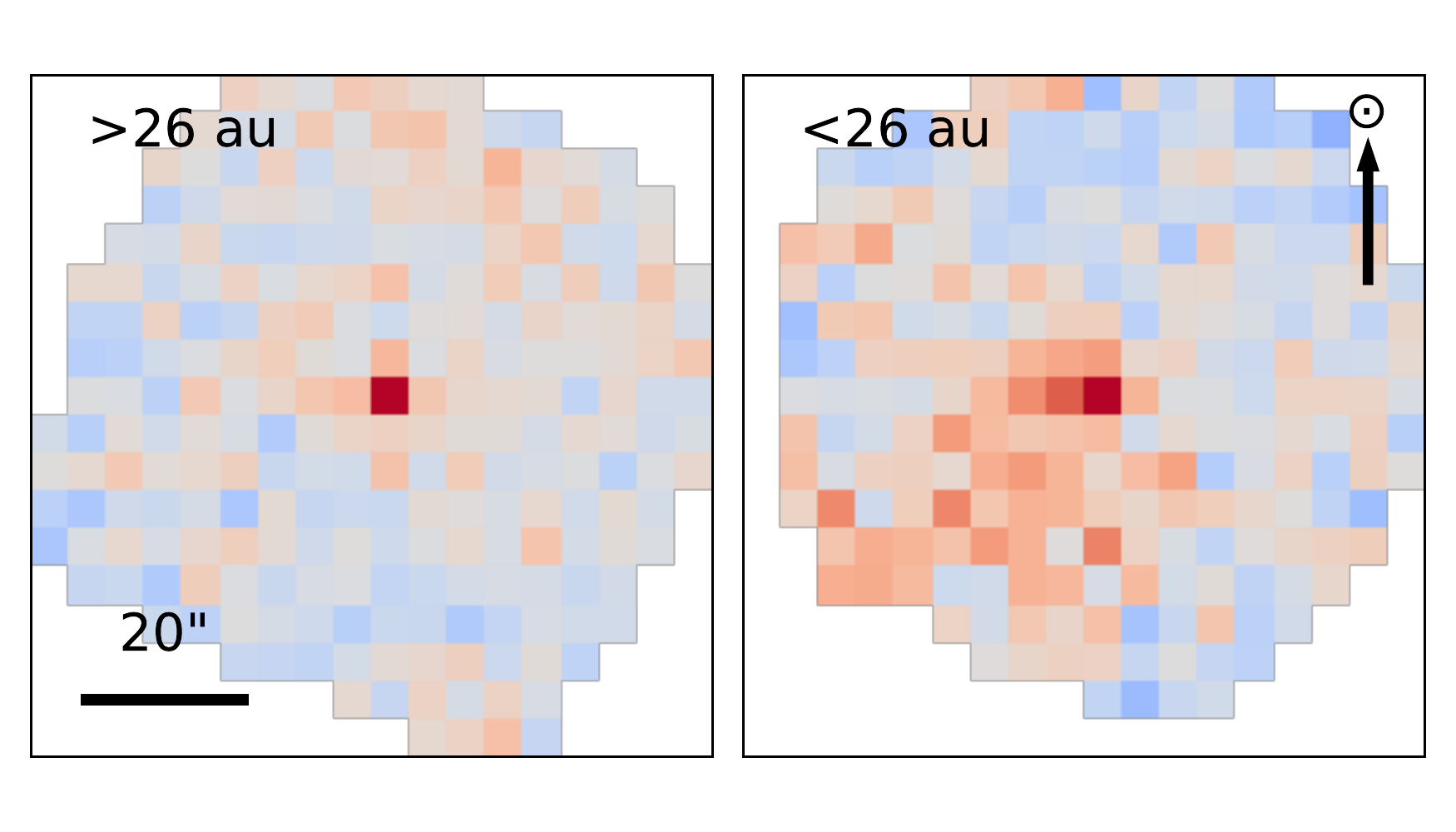}
  \caption{The averages of \des\ images of \bb\ taken exterior (left) and
    interior (right) to $r_h=26\au$ are displayed on the same angular
    and flux scales.  These are scene-modeling residual images after subtraction of
    the background model and a point-source model of the nucleus.  All
    images are scaled to $r$-band using solar colors, inverse-variance
    weighted, rotated such that the projected direction toward the Sun
    is vertical, and binned to
    $\approx4\farcs5$ pixel size.  The development of a tail or coma
    during the \des\ observations is apparent, but it is not precisely
    aligned with the anti-solar vector.}
  \label{im:binned}
\end{figure}

The curve of growth from the \des\ 2018 season
is plausibly consistent with
the measured \tess\ magnitude made in its $(5\times21)\arcsec$
square aperture
during the same season.  Post-\des\ imaging clearly shows exponential
  increase in coma brightness, which we will quantify below.

We  take a closer look at the structure 
and history of the coma
during the \des\ epochs using the
residual images produced by the scene-modeling photometry after
subtraction of the static sky background and the best-fit central
point source.  Each image is scaled to $r$-band assuming solar
colors---unfortunately we have insufficient $S/N$ to
meaningfully constrain the coma color. Residual artifacts from
defects, cosmic rays, and misregistration are masked.
Figure~\ref{im:binned} shows the inverse-variance-weighted average of these,
split between the first three seasons ($r_h>26\au$) and the last two
($r_h<26\au).$ Growth of a tail or asymmetric coma during this epoch
is apparent. 
An anti-solar tail would point downwards in this image stack; the
observed diffuse light is $\approx 40\degr$ away from anti-solar.

\begin{figure}
  \plottwo{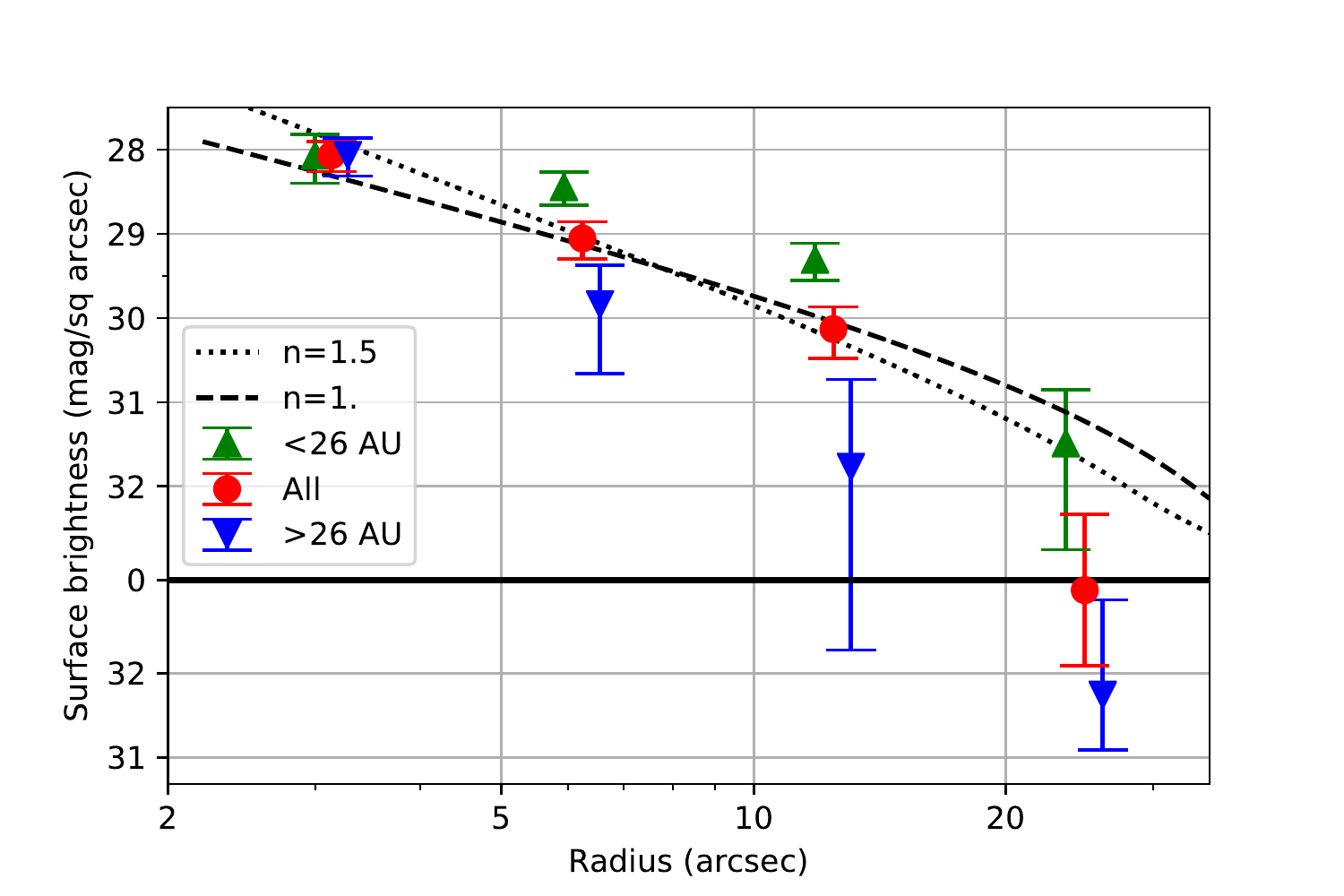}{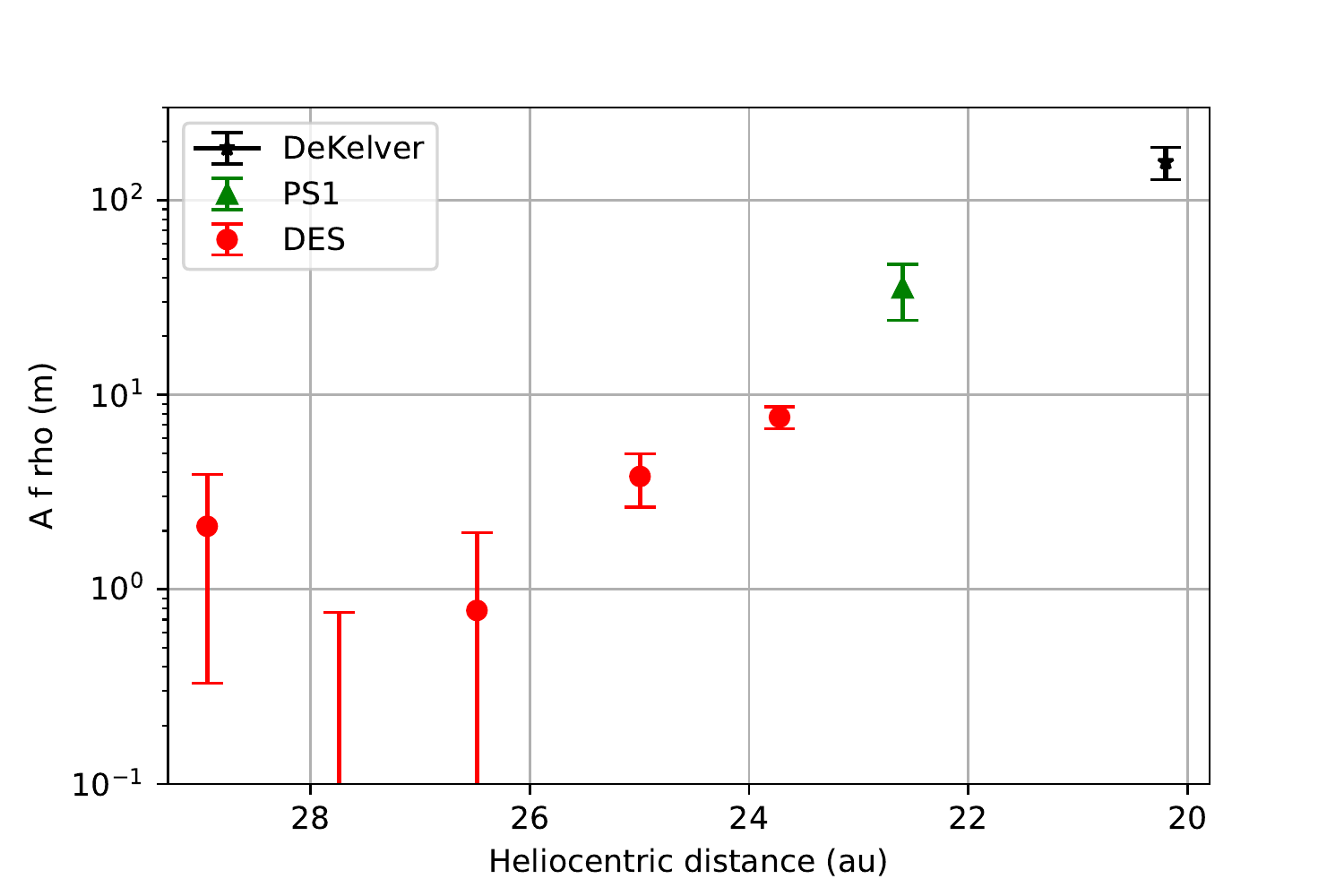}
  \caption{\textit{Left}: the surface brightness of the coma in annular bins as
    measured from \des\ images.  [Note the $y$ axis is logarithmic in flux
    (linear in mag) above 32~mag~arcsec$^{-2}$, linear in flux between
    this point and its negative-SB counterpart, and logarithmic in
    (negative) flux near the bottom.]
    The average over all \des\ exposures is shown, as well
    as split into data before 2017, ($29>r_h>26\au$) and
    in 2017--2018, ($26>r_h>23\au$).  The curves show models with
    surface brightness $I\propto\rho^{-n}$ for $n=1,$ as expected for
    a stationary coma, and $n=1.5$, as expected for a
    radiation-pressure-driven dust coma.  Either is consistent with
    the data.
    \textit{Right:} Under the
    stationary coma model, the inferred $\afrho$ is plotted against
    $r_h$, for each \des\ observing season, and for
    later observations with well-defined aperture magnitudes. The coma
    reflectivity grows exponentially }
  \label{im:descoma}
\end{figure}

More quantitative measures of the growth of coma are plotted in
Figures~\ref{im:descoma}.  
The left panel shows the
results in the observational space of surface brightness in annular
bins of radius.
The surface brightness $I$ scaling with radius $I\propto \rho^{-n}$ is
consistent either a ``stationary'' coma, $n=1,$ as expected if dust
particles move ballistically at fixed $v_d$ from the nucleus; or with
$n=1.5,$ as is suggested by models of radiation-pressure-dominated
escape \citep{JewittMeech87}.
We
are pleased to see that the coma is well measured even at surface
brightness below 30~mag~arcsec$^{-2}$, which generates
$<0.004e$/sec/pixel in the images, a tribute to the quality of the
image calibration and the background subtraction in the scene-modelling
method.

The standard measure of coma surface brightness is $\afrho$, where $A$
is the geometric albedo and $f$ is the filling factor of the
reflecting particles, conventionally measured as the average
$f(<\rho)$ interior to radius $\rho.$   For a stationary coma, this
quantity is invariant with the distances from the source, the sun, and
the observer \citep{ahearn_afrho, fink}.  We transform the surface
brightness into $\afrho$ via
\begin{align}
  I(\rho) & = \frac{Af(<\rho)}{2} \frac{L_\odot}{16\pi^2 r_h^2} \\
  \Rightarrow \qquad A\! f\! \rho & = 2 \times 10^{0.4(M_\odot-m_{\rm SB})} 4\pi\rho
                               \left(\frac{r_h}{10\au}\right)^2,
\end{align}
where $M_\odot$ is the absolute magnitude of the sun, and $m_{\rm SB}$
is the observed surface brightness per arcsec$^2.$  The right panel of
Figure~\ref{im:descoma} plots the $\afrho$ inferred from
  fitting a stationary coma model to each \des\ exposure at radii
  $\ge4\arcsec$ from the nucleus, and averaging over each
season's averaged observations.  This is plotted vs $r_h$, and we
include values taken from later observations.  From \citet{dekelver},
we take the uncertainty to be the span of $Af\rho$ values determined
at different radii.  For the \textit{PS1} observations, we apply a
generous $\pm30\%$ standard error.
The data exhibit an
exponential increase in the dust content of the coma,
growing $\approx2\times$ with each au reduction in $r_h.$  The
\tess\ data even exhibit this rate of brightening within the
duration of its 2020 observing (8\% per month).
At $r_h>26\au,$ the uncertainties are large enough to admit
a wide variety of behavior, \textit{e.g.} even
a constant coma surface brightness as might occur if \bb\ entered the
inner solar system with a gravitationally bound ``dirtmosphere'' of
particles accumulated through impacts over millions of years.

\section{Discussion}

\bb\ has uniquely high quality photometric data through the initial
growth period of its coma, with direct detections of coma out to
$r_h\approx26\au.$  We expect future work to produce detailed
thermal and dynamical modeling of the comet, but here we show that a
very simple model fits the observations well.

\subsection{Sublimating species}

For a single species with molecular mass $m_{\rm mol}$
sublimating into vacuum from a surface, the mass loss rate per unit
area $A$ is
\begin{align}
  \label{evaporation}
  \dot m \equiv \frac{\dot M}{A} & = P_{\rm sat} \sqrt{\frac{m_{\rm mol}}{2\pi k T}} \\
\label{clausius}
  & \propto e^{-\Delta H/RT} \sqrt{\frac{m_{\rm mol}}{2\pi k T}} \\
     \Rightarrow \qquad \log\dot M\sqrt{T}  &= \textrm{const} - \frac{\Delta
                 H}{RT}.
\label{logM}
\end{align}
where $P_{\rm sat}$ is the saturation vapor pressure, and we use the
Clausius-Clapeyron formula to express its dependence on temperature
$T$ in (\ref{clausius}).  $\Delta H$ is the enthalpy of sublimation
(which we assume varies little with $T$) and
$R$ is the ideal gas constant.

Under radiative equilibrium with negligible heat conduction with the
cometary interior and negligible heat loss to sublimation, a section
of the surface attains temperature $T$ with
\begin{align}
  \epsilon \sigma T^4 & =\frac{L_\odot}{4 \pi r_h^2} (1-p) \langle\cos\theta\rangle \\
\label{radeq2}
  \Rightarrow \qquad T & = \left[ \frac{(1-p)\langle\cos\theta\rangle}{\epsilon}\right]^{1/4}
                         \left(\frac{L_\odot}{4 \pi \sigma 
                         (1\au)^2}\right)^{1/4}
                         \left(\frac{r_h}{1\au}\right)^{-1/2} \\
                 & = \eta \times \left(397\,\textrm{K}\right)
                   \left(\frac{r_h}{1\au}\right)^{-1/2}.
                   \label{radeq}
\end{align}
In these equations, $\sigma$ is the Stefan-Boltzmann constant, $p$ is
the Bond albedo of the surface (nominally 0.04),
and $\epsilon$ is the infrared emissivity (nominally 0.9).
With $\theta$ as the angle between illumination and the
normal, the average $\langle\cos\theta\rangle$  over the thermal time
scale for the warmest part of the comet (which will dominate the
sublimation rate at low $T$) will be between 1 (at the subsolar point
for short thermal time constant, or for a pole-on rotator) and $1/\pi$
for the equator of a orthogonal rotator with long time constant.  We
bundle all of these physical/geometric constants in the first term of
(\ref{radeq2} into a factor $\eta,$ which is nominally close to
unity but could be as low as $\approx0.75.$

The third part of the simple model is to relate the coma brightness
$\afrho$ to the sublimation rate.  If the scattering is dominated by
solid particles with albedo $p_d$, radius $a_d$ and near-spherical,
geometric cross-section, density $\rho_d$, production rate $\dot M_d$,
and velocity $v_d$, then
\begin{equation}
    \afrho = \frac{3 p_d \dot M_d}{8 \pi a_d \rho_d v_d}.
\label{eq:massloss}
\end{equation}
If we assume that $p_d, a_d, \rho_d,$ and the dust-to-gas ratio
$\chi=\dot M_d / \dot M$ are independent of heliocentric distance over
the 20--30~au range, we obtain a scaling
\begin{equation}
  \afrho \propto \dot M / v_d.
  \label{eq:massloss2}
\end{equation}
Note that this proportionality does not require the
geometric-scattering limit to hold, only that scattering per unit mass
of dust is time-invariant.
Combining this with Eq.~(\ref{logM}) yields
\begin{equation}
\log \left( \afrho\, v_dT^{1/2}\right) = \textrm{const} - \frac{\Delta
                 H}{RT}.
   \label{eq:logafrho}
\end{equation}
One working assumption for $v_d$ is that it will scale with the
thermal velocity, i.e. $\propto T^{1/2}.$  A stronger dependence would
be expected if the dust velocity is driven by radiation pressure: $v_d
\propto r_h^{-2} \propto T^4.$ 

\begin{figure}
  \plottwo{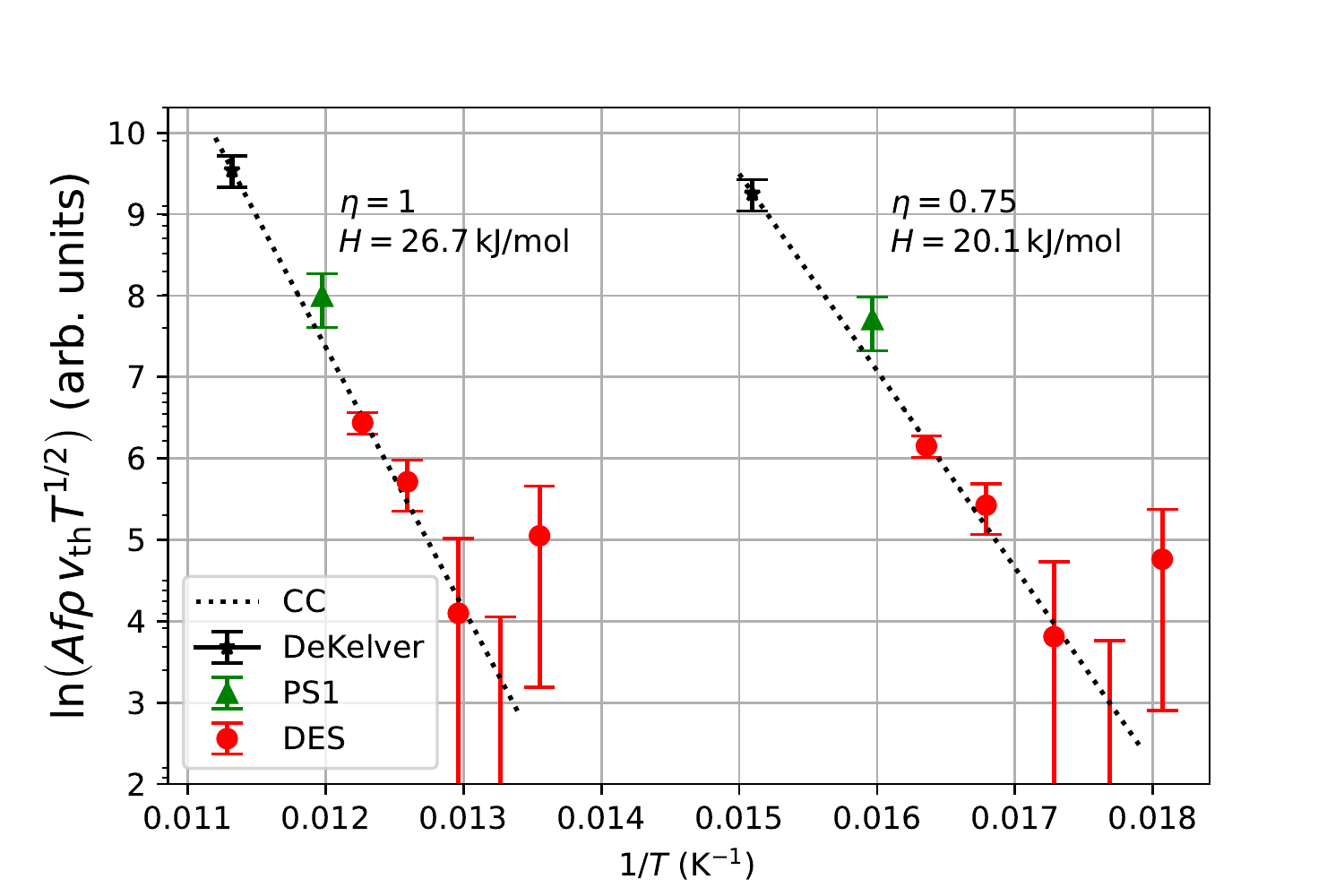}{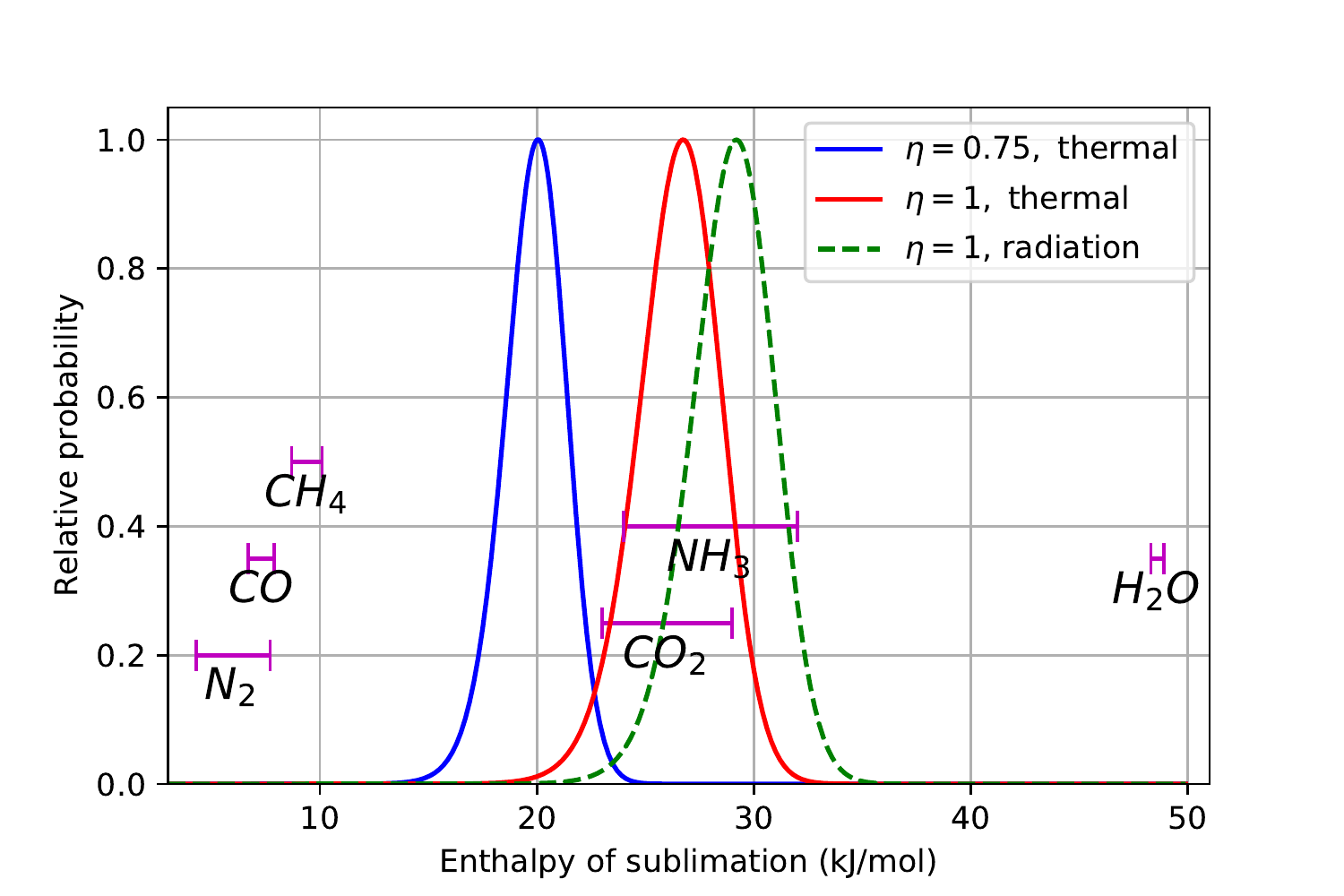}
  \caption{\textit{Left:} Following Eq.~\ref{eq:logafrho} and assuming
    radiative equilibrium temperatures, we plot the log of $\afrho\, v_{\rm th}
    \sqrt{T}$ vs $1/T,$ which should yield a straight line if the
    coma's scattering strength is proportional to the sublimation rate
    of a single species following the Clausius-Clapeyron (CC)
    relation.  For either the fast- or slow-rotator bounds 
    on radiative equilibrium ($\eta=0.75,1$), the data are well fit by such a form.
    \textit{Right:} Relative probability of the enthalpy of
    sublimation $\Delta H$ in a fit of Eq.~\ref{eq:logafrho} to the measured
    values of $\afrho,$ marginalizing over the scaling constant.  The
    central red curve gives the nominal case, with $\eta=1$ and a dust velocity scaling
    with the thermal velocity.  The left blue curve assumes a fast-rotating
    limit ($\eta=0.75$) and the right dashed curve assumed dust
    velocities scaling with radiation pressure ($v_d\propto T^4$).  The enthalpies of
    sublimation of potential cometary volatiles are marked: the data
    strongly favor $CO_2$ or $NH_3$ as the driver of \bb's mass loss
    to date.}
  \label{im:massloss}
\end{figure}

The left panel of Figure~\ref{im:massloss} plots $\afrho\, v_d
T^{1/2}$ on the (logarithmic) $y$ axis, vs $1/T$ on the $x$ axis,
assuming radiative equilibrium for $T$ and $v_d\propto v_{\rm th}.$ As
per Eq.~\ref{eq:logafrho}, the data should follow a line with slope
$-\Delta H/R$ on this plot for sublimation of a single species.  The data are
seen to be consistent with this model, with a value of $\Delta H$ that
depends on whether we take the fast- or slow-rotator limit for $\eta$
in the radiative equilibrium formula.

The right panel of Figure~\ref{im:massloss} plots the result of a
quantitative fit of Eq.~\ref{eq:logafrho} to the $\afrho$
measurements as function of $\Delta H$, marginalizing over the
unknown constant.  In the nominal (red) case, the inferred enthalpy of
sublimation is fully consistent with the lab-measured values for
$NH_3$ and/or $CO_2,$ and clearly inconsistent with the more volatile
species $N_2, CH_4,$ and $CO$.  As is well known, $H_2O$ is ruled out
as a volatile at these distances as well. The values of
$\Delta H$ are taken from \citet{luna,Feistel}.  The green dashed curve
changes the scaling of $v_d$ vs $T$ from thermal to radiation-pressure
laws, and does not change the conclusion.  Moving to the fast-rotator
value $\eta=0.75$ yields proportionately lower $\Delta H$ values, still
better associated with $CO_2$ and $NH_3.$

The coma growth rate is thus strongly suggestive of activity powered
by $CO_2$ and/or $NH_3$ sublimation at $20<r_h<25\au$, with slow and/or
pole-on rotation somewhat favored to yield higher peak surface
temperatures. Across the range of potential surface temperatures, the
mass loss rate per square meter from pure 
$CO_2$ is $>100\times$ larger than that from a pure $NH_3$ ice
surface, so the former is more likely to drive activity.

At $r_h>25\au,$ the coma is too weak to be well
measured in the data.  It is possible that more volatile species
($N_2, CH_4,$ or $CO$) could have dominated sublimation at these
times.  These latter species, however, have orders of magnitude higher
vapor pressure (and specific sublimation rates) than $CO_2$ and $NH_3$
do at $T=78$~K, the $\eta=1$ subsolar temperature for $r_h=25\au.$ Any surface
abundance in their pure ice forms on the surface of \bb\ must be very
low
relative to the $CO_2$/$NH_3$ that appear to dominate sublimation at
$r_h\le25\au.$   Perhaps these more-volatile species were heavily
sublimated from \bb\ during its previous perihelion passage
to $\approx18\au,$, leaving behind a crust that is largely depleted
in these most volatile species. Or perhaps this depletion is a
  remnant of the thermal environment at the original location of
  formation of BB.

We have not investigated activity powered by phase changes or
annealing of ices.

\subsection{Dust production}
The dynamics of dust production on \bb\ are made more interesting
by the fact that the escape velocity of $\approx60$~m/s is above or
comparable to the dust velocities estimated for other comets,
\textit{e.g.} $<50$~m/s for Comet Boattini \citep{boattini} or
$\approx4$~m/s for C/2017 K2 \citep{Jewitt2019}.
The low sublimation rates and pressures for \bb\ at
$r_h>22\au$ may limit the size of particles that can be
raised and attain escape by the gas, and/or the coma may be dominated
by small grains that are driven to escape by radiation pressure.
The coma is observed to be
asymmetric in its 2021 observations, in the 2018 \des\ data
(Figure~\ref{im:binned}), and  potentially in the
\tess\ observations \citep{TESS}, implicating radiation pressure or
tidal escape mechanisms for the dust. High-quality imaging of the coma as soon as
possible would be of great interest in constraining the dust size and
dynamics. 

If a fraction $f_{\rm active}$ of the surface of \bb\ is sublimating
$CO_2$ at the rates described in Eq.~\ref{evaporation}, then the gas mass
loss rate from \bb\ rises from $400 f_{\rm active}$ to $7\times10^4 f_{\rm active}$~kg/s 
when closing from 26 to 20~au.  This assumes $\eta\approx1$ (slow
rotator), and the vapor pressure cited by \citet{Fray}.  A non-porous,
pure-$CO_2$ ice patch at the subsolar point would be eroded by
0.1~mm/yr at 26~au, increasing to 2~cm/yr at its current
$r_h\approx20\au.$ 

Adopting the simplistic model of Eq.~\ref{eq:massloss} and
a uniform streaming velocity
$v_d=f_vv_{\rm th},$ where $v_{\rm th}$ is the RMS 1d thermal
velocity of a $CO_2$ molecule, then we can solve for the dust particle radius
as 
\begin{align}
  a_d & = \sqrt{\frac{9}{8\pi}} \frac{p_d \chi f_{\rm active}} {f_v}
        \frac{P_{\rm sat} a_{BB}^2 m_{\rm mol}}{\rho_d [A\!f\!\rho] kT} \\
   & \approx 
  \left(\frac{p\chi}{0.04}\right)\left(\frac{0.1}{f_v}\right)
  \left(\frac{f_{\rm active}}{0.1}\right) \times
  10\,\mu{\rm m}.
\label{dustsize}
\end{align}
We take
$a_{\rm BB}$ as the nominal radius of the comet, $m_{\rm mol}$
is the mass of a $CO_2$ molecule, and albedos of $p=0.04$ for both comet and
its debris, and the observed values of $\afrho$ to obtain (\ref{dustsize}).
The nominal values of $\chi, f_{\rm active},$ and $f_v$
  are ill-supported order-of-magnitude estimates. The implied
  dust particle radius  may even be over-estimated as the
  assumption of $f_{\rm active}=0.1$ of the surface being $CO_2$ ice
  near the subsolar temperature may be an over-estimate.
 It nonetheless suggests that the coma has small
particles, indeed small enough that the assumption of geometric
scattering cross-section needs to be relaxed.  Small dust particles
might be expected because of the anemic gas flow density and the 
importance of radiation pressure in overcoming \bb's gravity.
Clearly a more detailed model of the
dust dynamics would be of interest.

\subsection{Comparison to other distant LPCs}
Opportunities to study incoming comets at $r_h>20\au$ have been rare
enough that there is no definition of ``typical'' behavior.  It is
already clear that the behavior is diverse.  It is good to keep in
mind that selection biases will favor the discovery at large distances
of comets that are unusually active or, like \bb, unusually large.

Comet C/2017~K2 (PanSTARRS), (hereafter ``K2'') has similar aphelion
and inclination to \bb, but \citet{krolikowska2018} report it to have a
97\% chance of having passed within $r_h<10\au$ on its previous
passage. It was discovered at $r_h=16\au$ but precovery images at
$r_h=24\au$ also display coma \citep{Hui18}.

K2 is much smaller than
\bb, with estimated nuclear radius $<9$~km \citep{JewittK2HST}.
These authors infer K2's
coma to be comprised of mm-scale particles moving at speeds of
$\approx4$~m/s---which would probably not escape \bb---and estimate that K2 has
been expelling such particles at a relatively steady and isotropic rate since
$r_h\approx30\au.$ This is in stark contrast to \bb's exponentially
growing scattering cross-section in the 20--30~au range.

Comet C/2010 U3 (Boattini) was discovered inbound at $r_H=18.4\au$ and
precovered on earlier images at $r_h=25.8$ and 24.6~au, with visible coma
\citep{boattini}. Like K2, Boattini is thought to have had its
previous perihelion at $<10\au$, yet its coma behavior is quite
distinct from K2's, showing intermittent outbursts and a tail inferred to be
composed of much smaller particles than K2's.

K2, Boattini, and \bb\ display very diverse activity patterns at
$r_h>15\au$.  K2 has a steady, large-particle coma, Boattini exhibits
outbursts and a tail, while \bb\ undergoes exponential growth in
cross-section over this period that is consistent with simple
sublimation thermodynamics of carbon dioxide and/or ammonia.    No generalized
pattern is apparent yet,
aside from the existence of activity in some form out to
$\approx30\au,$ and even this ubiquity can be favored by selection
effects.

\section{Summary}
\unbb\ is arguably the largest and
most pristine comet ever discovered.  Assuming a typical albedo, its
diameter of $\approx150$~km implies a mass $10\times$ larger than
Hale-Bopp, and capable of gravitationally binding many of the larger
particles ejected from other comets.  The object is at present $20\au$
from the sun, and its 
previous perihelion was likely at $\gtrsim18\au,$ so this may be the
only comet ever measured before any approach to $r_h<10\au.$   Its
nucleus has nearly gray reflectivity, in common with (or slightly bluer than)
previously observed objects with Oort-cloud origins.

Variation in the absolute magnitude of the nucleus is strongly
detected, and consistent with a $\pm0.20$~mag sinusoidal light curve.
The data are too sparse to actually derive a light curve.

We are able to detect the presence of activity in \des\ images starting with the 2017
season, at $r_h\approx25\au,$ which grows exponentially with approach
to $20\au.$  The rate of growth is consistent with sublimation of a
species with enthalpy of sublimation near the 26~kJ/mol of $CO_2,$ (or
$NH_3$). 
The coma measurements at $r_h>25\au$ have too low $S/N$ to
  characterize the behavior, so these earlier phases of activity could
  e.g. be dominated by other species'
sublimation.
\bb\ is thus an outlier among the (notoriously unpredictable)
population of comets in that its onset of activity follows simple
sublimation thermodynamics to date, \textit{i.e.} it is (so far) a
``spherical cow.'' Perhaps this behavior is related to the fact that
it is also an outlier in size and in its relatively
uneventful past thermal history. 

It is usually a losing proposition to speculate on the
future behavior of comets, even one such as \bb\ whose activity has
followed a simple model to date.  Ignoring this warning, we
  can make an estimate of \bb's 
brightening if its scattering cross-section continues to grow in
proportion to the $CO_2$ sublimation rate in radiative equilibrium as
it reaches its 11~au perihelion in a decade.  The $CO_2$ sublimation
rate will grow to a level at which most of the incident solar flux is
turned to sublimation enthalpy.  The mass loss rate then scales
with the fraction of the surface material composed of $CO_2$
ice near the subsolar point.  If this is 10\%, then the sublimation
rate will be $\gtrsim200\times$ above its value in June 2021.
Combined with the $r_h^{-4}$ brightening, \bb\ would be 8.5~mag
brighter in apparent magnitude at perihelion than the magnitude
$G\approx17.5$ currently report in large apertures
\citep{dekelver}---i.e. $G\approx9,$ a bit fainter than Titan.  If a
water-ice crust forms and blocks $CO_2$ sublimation, the coma would
be suppressed. If the $CO_2$
sublimation spreads across more of the comet's surface, it could be
substantially brighter.
It will be an impressive telescopic target, and its large surface area
may generate a substantial $CO_2$-powered coma and tail despite
remaining far outside the water-ice line.

The catalog of distant incoming Oort comets is likely to grow rapidly
in the next decade, as the \textit{Vera C. Rubin Observatory Legacy
  Survey of Space and Time} will easily detect and track any
object of half \bb's size that comes within $r_h\lesssim40\au$ in the next
decade, even those with no activity, obtaining hundreds of exposures
in multiple bands for each.

\begin{acknowledgements}

We thank Gonzalo Tancredi and Jim Annis for careful reviews of this
work before submission.
University of Pennsylvania authors have been supported in this work by
grants AST-1515804 and AST-2009210 from the National Science
Foundation, and grant DE-SC0007901 from the Department of Energy.  

Pan-STARRS is supported by the National Aeronautics and Space
Administration under Grant No. 80NSSC18K0971 issued through the SSO
Near Earth Object Observations Program.

This research has made use of the services of the ESO Science Archive
Facility. Based on observations collected at the European Southern
Observatory under ESO program 179.A-2004(C).  This work also makes use
of public data from the Canada-France-Hawaii Telescope Science Archive.

This paper includes data collected by the \tess\ mission, which are
publicly available from the Mikulski Archive for Space Telescopes
(MAST). Funding for the \tess\ mission is provided by NASA’s Science
Mission directorate.

Funding for the DES Projects has been provided by the U.S. Department of Energy, the U.S. National Science Foundation, the Ministry of Science and Education of Spain, 
the Science and Technology Facilities Council of the United Kingdom, the Higher Education Funding Council for England, the National Center for Supercomputing 
Applications at the University of Illinois at Urbana-Champaign, the Kavli Institute of Cosmological Physics at the University of Chicago, 
the Center for Cosmology and Astro-Particle Physics at the Ohio State University,
the Mitchell Institute for Fundamental Physics and Astronomy at Texas A\&M University, Financiadora de Estudos e Projetos, 
Funda{\c c}{\~a}o Carlos Chagas Filho de Amparo {\`a} Pesquisa do Estado do Rio de Janeiro, Conselho Nacional de Desenvolvimento Cient{\'i}fico e Tecnol{\'o}gico and 
the Minist{\'e}rio da Ci{\^e}ncia, Tecnologia e Inova{\c c}{\~a}o, the Deutsche Forschungsgemeinschaft and the Collaborating Institutions in the Dark Energy Survey. 

The Collaborating Institutions are Argonne National Laboratory, the University of California at Santa Cruz, the University of Cambridge, Centro de Investigaciones Energ{\'e}ticas, 
Medioambientales y Tecnol{\'o}gicas-Madrid, the University of Chicago, University College London, the DES-Brazil Consortium, the University of Edinburgh, 
the Eidgen{\"o}ssische Technische Hochschule (ETH) Z{\"u}rich, 
Fermi National Accelerator Laboratory, the University of Illinois at Urbana-Champaign, the Institut de Ci{\`e}ncies de l'Espai (IEEC/CSIC), 
the Institut de F{\'i}sica d'Altes Energies, Lawrence Berkeley National Laboratory, the Ludwig-Maximilians Universit{\"a}t M{\"u}nchen and the associated Excellence Cluster Universe, 
the University of Michigan, NSF's NOIRLab, the University of Nottingham, The Ohio State University, the University of Pennsylvania, the University of Portsmouth, 
SLAC National Accelerator Laboratory, Stanford University, the University of Sussex, Texas A\&M University, and the OzDES Membership Consortium.

Based in part on observations at Cerro Tololo Inter-American Observatory at NSF's NOIRLab (NOIRLab Prop. ID 2012B-0001; PI: J. Frieman), which is managed by the Association of Universities for Research in Astronomy (AURA) under a cooperative agreement with the National Science Foundation.

The DES data management system is supported by the National Science Foundation under Grant Numbers AST-1138766 and AST-1536171.
The DES participants from Spanish institutions are partially supported by MICINN under grants ESP2017-89838, PGC2018-094773, PGC2018-102021, SEV-2016-0588, SEV-2016-0597, and MDM-2015-0509, some of which include ERDF funds from the European Union. IFAE is partially funded by the CERCA program of the Generalitat de Catalunya.
Research leading to these results has received funding from the European Research
Council under the European Union's Seventh Framework Program (FP7/2007-2013) including ERC grant agreements 240672, 291329, and 306478.
We  acknowledge support from the Brazilian Instituto Nacional de Ci\^encia
e Tecnologia (INCT) do e-Universo (CNPq grant 465376/2014-2).

This manuscript has been authored by Fermi Research Alliance, LLC under Contract No. DE-AC02-07CH11359 with the U.S. Department of Energy, Office of Science, Office of High Energy Physics. The United States Government retains and the publisher, by accepting the article for publication, acknowledges that the United States Government retains a non-exclusive, paid-up, irrevocable, world-wide license to publish or reproduce the published form of this manuscript, or allow others to do so, for United States Government purposes.


\end{acknowledgements}
\newpage
\startlongtable
\begin{deluxetable}{llllclCcc}
  \tablecaption{Observational data for C/2014 UN$_{271}$}
  \label{obstable}
\tablehead{\colhead{UTC} & \colhead{RA} & \colhead{Dec} & \colhead{Error} & \colhead{Source} & \colhead{Mag} & \colhead{Band} & \colhead{Helio. dist.} & \colhead{Geo. dist.} \\ 
 & & & & & & & \colhead{(au)} & \colhead{(au)} }
\startdata
2010-11-15.238223 & 01:19:36.7532 & -30:42:27.260 & 0\fs105 & VISTA & 22.47$\pm$0.26 & z &                  34.07 & 33.52 \\
2014-08-14.599780 & 01:42:23.9960 & -35:36:37.600 & 0\fs078 & PS1 & 23.10$\pm$0.20$\ast$ & i &              29.28 & 28.74 \\
2014-08-28.601410 & 01:41:25.1473 & -36:02:11.803 & 0\fs04 & CFHT & 22.39$\pm$0.09 & r &                    29.23 & 28.59 \\
2014-08-28.606693 & 01:41:25.1180 & -36:02:12.397 & 0\fs09 & CFHT & 23.05$\pm$0.08 & g &                    29.23 & 28.59 \\
2014-08-28.609338 & 01:41:25.1015 & -36:02:12.632 & 0\fs06 & CFHT & 22.14$\pm$0.17 & i &                    29.23 & 28.59 \\
2014-08-28.612016 & 01:41:25.0891 & -36:02:13.023 & 0\fs04 & CFHT & 22.69$\pm$0.12 & r &                    29.23 & 28.59 \\
2014-10-20.294348 & 01:34:35.0187 & -37:14:46.144 & 0\fs068 & DES & 22.62$\pm$0.10 & r &                    29.05 & 28.36 \\
2014-11-04.122035 & 01:32:21.0290 & -37:23:21.138 & 0\fs050 & DES & 22.09$\pm$0.12 & z &                    28.99 & 28.41 \\
2014-11-14.258286 & \nodata & \nodata & \nodata & DES & 21.93$\pm$0.46 & Y &                                28.96 & 28.46 \\
2014-11-15.251906 & 01:30:47.0392 & -37:25:39.196 & 0\fs047 & DES & 22.32$\pm$0.16 & z &                    28.95 & 28.47 \\
2014-11-18.235595 & 01:30:23.4255 & -37:25:40.010 & 0\fs040 & DES & 22.52$\pm$0.05 & r &                    28.94 & 28.49 \\
2014-11-18.238347 & 01:30:23.4067 & -37:25:39.984 & 0\fs041 & DES & 22.22$\pm$0.07 & i &                    28.94 & 28.49 \\
2014-11-27.221480 & 01:29:17.5832 & -37:24:12.514 & 0\fs113 & DES & \nodata & g &                                 28.91 & 28.55 \\
2014-12-11.115676 & 01:27:54.8295 & -37:17:49.638 & 0\fs084 & DES & 22.43$\pm$0.09 & i &                    28.86 & 28.67 \\
2014-12-11.144452 & \nodata & \nodata & \nodata & DES & 22.57$\pm$0.62 & Y &                                28.86 & 28.67 \\
2015-01-09.107010 & 01:26:35.3245 & -36:51:38.750 & 0\fs130 & DES & 22.65$\pm$0.30 & z &                    28.76 & 28.94 \\
2015-08-11.612980 & 01:47:05.4300 & -37:26:12.860 & 0\fs102 & PS1 & 23.33$\pm$0.22$\ast$ & i &              27.99 & 27.49 \\
2015-08-17.357470 & 01:46:47.5049 & -37:37:39.814 & 0\fs053 & DES & 22.64$\pm$0.08 & r &                    27.97 & 27.42 \\
2015-08-17.358826 & 01:46:47.5170 & -37:37:39.886 & 0\fs080 & DES & 22.99$\pm$0.09 & g &                    27.97 & 27.42 \\
2015-08-24.344520 & 01:46:18.6053 & -37:51:28.614 & 0\fs060 & DES & 23.12$\pm$0.09 & g &                    27.94 & 27.35 \\
2015-08-24.347311 & \nodata & \nodata & \nodata & DES & 22.53$\pm$0.07 & r &                                27.94 & 27.35 \\
2015-08-24.348700 & \nodata & \nodata & \nodata & DES & 22.38$\pm$0.08 & i &                                27.94 & 27.35 \\
2015-09-01.306625 & \nodata & \nodata & \nodata & DES & 22.46$\pm$0.54 & Y &                                27.92 & 27.27 \\
2015-09-02.377866 & 01:45:30.1964 & -38:08:52.304 & 0\fs085 & DES & 21.94$\pm$0.28 & Y &                    27.91 & 27.26 \\
2015-09-13.390509 & 01:44:15.8625 & -38:28:51.622 & 0\fs073 & DES & 22.47$\pm$0.10 & i &                    27.87 & 27.17 \\
2015-10-06.274354 & 01:40:59.7742 & -39:03:19.836 & 0\fs147 & DES & 22.18$\pm$0.17 & z &                    27.79 & 27.08 \\
2015-11-20.228513 & 01:33:55.0781 & -39:29:32.567 & 0\fs067 & DES & 22.18$\pm$0.14 & z &                    27.63 & 27.21 \\
2015-11-20.235561 & 01:33:55.0098 & -39:29:32.478 & 0\fs058 & DES & 22.50$\pm$0.11 & i &                    27.63 & 27.21 \\
2016-01-11.093863 & 01:29:57.7382 & -38:50:26.056 & 0\fs044 & DES & 22.45$\pm$0.06 & r &                    27.44 & 27.64 \\
2016-01-11.095236 & 01:29:57.7285 & -38:50:25.964 & 0\fs055 & DES & 23.08$\pm$0.09 & g &                    27.44 & 27.64 \\
2016-08-09.585360 & 01:52:24.3560 & -39:29:37.720 & 0\fs090 & PS1 & 23.53$\pm$0.18$\ast$ & w &              26.68 & 26.21 \\
2016-08-09.595282 & 01:52:24.3230 & -39:29:39.130 & 0\fs129 & PS1 & 23.45$\pm$0.22$\ast$ & w &              26.68 & 26.21 \\
2016-10-01.297615 & \nodata & \nodata & \nodata & DES & 22.10$\pm$0.07 & i &                                26.49 & 25.80 \\
2016-10-01.298994 & 01:46:40.1803 & -41:11:31.779 & 0\fs043 & DES & 22.16$\pm$0.05 & r &                    26.49 & 25.80 \\
2016-10-01.300363 & 01:46:40.1727 & -41:11:32.002 & 0\fs069 & DES & 22.63$\pm$0.05 & g &                    26.49 & 25.80 \\
2016-10-03.314076 & \nodata & \nodata & \nodata & DES & 22.43$\pm$0.10 & i &                                26.48 & 25.80 \\
2016-10-03.315455 & 01:46:20.1770 & -41:14:24.527 & 0\fs038 & DES & 22.36$\pm$0.05 & r &                    26.48 & 25.80 \\
2016-10-03.316819 & 01:46:20.1667 & -41:14:24.592 & 0\fs050 & DES & 22.81$\pm$0.07 & g &                    26.48 & 25.80 \\
2017-08-13.571449 & 01:58:27.6990 & -41:56:20.770 & 0\fs114 & PS1 & 22.52$\pm$0.18$\ast$ & i &              25.34 & 24.86 \\
2017-08-14.585122 & 01:58:24.7020 & -41:58:43.760 & 0\fs054 & PS1 & 21.70$\pm$0.09$\ast$ & i &              25.34 & 24.85 \\
2017-09-08.554246 & 01:56:09.8760$\ast$ & -42:55:16.230$\ast$ & 0\fs102 & PS1 & 22.00$\pm$0.20$\ast$ &  i & 25.24 & 24.62 \\
2017-09-30.421878 & 01:52:49.6350 & -43:36:32.720 & 0\fs112 & PS1 & 21.96$\pm$0.15$\ast$ & i &              25.16 & 24.51 \\
2017-10-15.272083 & 01:50:05.7460 & -43:57:07.295 & 0\fs056 & DES & 22.29$\pm$0.05 & g &                    25.11 & 24.49 \\
2017-10-15.337769 & 01:50:04.9880 & -43:57:11.797 & 0\fs063 & DES & 21.42$\pm$0.08 & z &                    25.11 & 24.49 \\
2017-10-15.339148 & 01:50:04.9697 & -43:57:11.890 & 0\fs031 & DES & 21.69$\pm$0.04 & r &                    25.11 & 24.49 \\
2017-10-29.358251 & 01:47:22.7320 & -44:09:45.240 & 0\fs145 & PS1 & 21.54$\pm$0.20$\ast$ & i &              25.06 & 24.52 \\
2017-11-06.330497 & 01:45:51.8610 & -44:13:42.320 & 0\fs131 & PS1 & 22.08$\pm$0.21$\ast$ & i &              25.03 & 24.54 \\
2017-12-11.250030 & 01:40:25.1950 & -44:04:24.940 & 0\fs127 & PS1 & 22.01$\pm$0.15$\ast$ & w &              24.90 & 24.76 \\
2017-12-15.180418 & 01:39:59.8371 & -44:00:56.398 & 0\fs029 & DES & 22.05$\pm$0.04 & r &                    24.89 & 24.79 \\
2017-12-15.181814 & 01:39:59.8244 & -44:00:56.379 & 0\fs049 & DES & 22.58$\pm$0.06 & g &                    24.89 & 24.79 \\
2017-12-25.149228 & \nodata & \nodata & \nodata & DES & 21.87$\pm$0.29 & Y &                                24.85 & 24.87 \\
2018-09-10.355826 & \nodata & \nodata & \nodata & DES & 22.06$\pm$0.03 & g &                                23.90 & 23.31 \\
2018-10-04 & \nodata & \nodata & \nodata & TESS & 20.29$\pm$0.15 & T &                                      23.82 & 23.21 \\
2018-10-21.243368 & 01:55:54.0391 & -46:47:14.935 & 0\fs075 & DES & 21.60$\pm$0.17 & Y &                    23.75 & 23.20 \\
2018-10-27.178589 & 01:54:38.2037 & -46:52:42.842 & 0\fs041 & DES & 21.59$\pm$0.07 & z &                    23.73 & 23.21 \\
2018-11-08.235135 & 01:52:05.9144 & -46:59:29.069 & 0\fs022 & DES & 21.71$\pm$0.03 & r &                    23.69 & 23.25 \\
2018-11-08.236509 & 01:52:05.8988 & -46:59:29.102 & 0\fs026 & DES & 21.51$\pm$0.04 & i &                    23.69 & 23.25 \\
2018-11-08.237892 & 01:52:05.8841 & -46:59:29.051 & 0\fs032 & DES & 22.23$\pm$0.04 & g &                    23.69 & 23.25 \\
2019-08-19.598297 & 02:14:55.2390 & -47:31:52.370 & 0\fs109 & PS1 & \nodata & i &                    22.64 & 22.18 \\
2019-08-29.569869 & 02:14:08.3560 & -47:59:21.660 & 0\fs188 & PS1 & \nodata & w &                   22.61 & 22.10 \\
2019-08-29.585234 & 02:14:08.2450 & -47:59:24.350 & 0\fs149 & PS1 & \nodata & w &                   22.61 & 22.10 \\
2019-08-29.592922 & 02:14:08.2250 & -47:59:25.260 & 0\fs144 & PS1 & \nodata & w &                    22.61 & 22.10 \\
2020-09-21 & \nodata & \nodata & \nodata & TESS & 18.24$\pm$0.15 & T &                                      21.18 & 20.67 
\enddata
\tablecomments{Data marked with $\ast$ are considered unreliable and not used in the analyses. Magnitudes for 2019 PS1 observations
are aperture-dependent, so not tabulated here.}
\end{deluxetable}

\bibliography{references}
\bibliographystyle{aasjournal}

\allauthors

\end{document}